\documentclass[sigplan,screen]{acmart}

\newcommand{\ignore}[1]{}
\newcommand{\red}[1]{\textcolor{red}{#1}}
\newcommand{\blue}[1]{\textcolor{blue}{#1}}
\newcommand{\green}[1]{\textcolor{ForestGreen}{#1}}

\usepackage{listings}
\usepackage{caption}
\usepackage{subcaption}
\usepackage{geometry}
\usepackage[capitalise,noabbrev,nameinlink]{cleveref}
\usepackage{array}
\usepackage{multirow}
\usepackage{makecell}
\usepackage{amsmath}
\usepackage{colortbl}
\usepackage{tikz}
\usetikzlibrary{calc}
\usepackage{soul}
\usepackage{xspace}
\usepackage{paralist}
\usepackage{enumitem}
\usepackage{ifthen}
\usepackage[title]{appendix}
\usepackage{etoolbox}
\usepackage{xcolor}
\patchcmd{\appendices}{\quad}{: }{}{}
\usepackage{balance} 

\usepackage[most]{tcolorbox}
\usepackage[frozencache=true,cachedir=minted-asplos24-paper-template]{minted}
\makeatletter
\tcbset{minted language/.store in=\kvtcb@minted@language}

\def\tcbuselistinglisting{%
  \toks@=\expandafter{\kvtcb@listingoptions}%
  \edef\tcb@temp{\noexpand\inputminted[\the\toks@]}%
  \tcb@temp{\kvtcb@minted@language}{\kvtcb@listingfile}%
}%
\makeatother
\newcommand{\compactpara}[1]{\smallskip\noindent\textbf{#1}}

\usepackage{pifont}
\newcommand{\cmark}{\ding{51}}%
\newcommand{\xmark}{\ding{55}}%

\definecolor{BrickRed}{HTML}{B6321C}
\definecolor{BurntOrange}{HTML}{F7921D}
\definecolor{GoldenRod}{HTML}{FFDF42}
\definecolor{ForestGreen}{HTML}{009B55}

\definecolor{keywordcolor}{rgb}{0.5,0.0,0.5}
\definecolor{stringcolor}{rgb}{0.0,0.5,0.0}
\definecolor{commentcolor}{rgb}{0.0,0.5,0.0}
\definecolor{functioncolor}{rgb}{0.0,0.0,0.8}

\lstdefinestyle{x86asm}{
    language=[x86masm]Assembler,
    basicstyle=\ttfamily\footnotesize,
    keywordstyle=\color{blue},
    commentstyle=\color{green!60!black},
    morecomment=[l][\color{green!60!black}]{\#},
    stringstyle=\color{red},
    numbers=left,
    numberstyle=\tiny\color{gray},
    stepnumber=1,
    numbersep=10pt,
    backgroundcolor=\color{gray!10},
    tabsize=4,
    showspaces=false,
    showstringspaces=false,
    frame=single,
    breaklines=true,
    morekeywords={mov, add, sub, mul, div, jmp, cmp, jne, je, call, ret, push, pop, int, xor, and, or, not,
    cmovb, cmovl, jmpq, xorb}
}

\lstdefinestyle{CStyle}{
    language=C,
    basicstyle=\ttfamily\footnotesize,
    keywordstyle=\color{blue},
    stringstyle=\color{red},
    commentstyle=\color{green!60!black},
    morecomment=[l][\color{magenta}]{\#},
    numbers=left,
    numberstyle=\tiny\color{gray},
    stepnumber=1,
    numbersep=10pt,
    backgroundcolor=\color{gray!10},
    tabsize=4,
    showspaces=false,
    showstringspaces=false,
    frame=single,
    breaklines=true,
    moredelim=**[is][\color{blue}]{@}{@},
    moredelim=**[is][\color{red}\ttfamily]{*}{*}
}

\lstdefinestyle{CStyle_NoLinenum}{
    language=C,
    basicstyle=\ttfamily\footnotesize,
    keywordstyle=\color{blue},
    stringstyle=\color{red},
    commentstyle=\color{green!60!black},
    morecomment=[l][\color{magenta}]{\#},
    backgroundcolor=\color{gray!10},
    tabsize=4,
    showspaces=false,
    showstringspaces=false,
    frame=single,
    breaklines=true,
    moredelim=**[is][\color{blue}]{@}{@},
    moredelim=**[is][\color{red}\ttfamily]{*}{*}
}

\newlist{rqenumerate}{enumerate}{1}
\setlist[rqenumerate,1]{label=\textbf{RQ\arabic*}, before=\setcounter{enumi}{1}}

\newcommand{\Gemfuzzer}{AMuLeT}
\newcommand{\tool}{\Gemfuzzer}
\newcommand{\naive}{\Gemfuzzer{}-Naive}
\newcommand{\opt}{\Gemfuzzer{}-Opt}
\newcommand{\utrace}{$\mu$arch trace\xspace}
\newcommand{\uarch}{$\mu$arch}

\newboolean{showcomments}

\setboolean{showcomments}{false}
\setboolean{showcomments}{true}

\newcommand{\rev}[1]{{\color{black}#1}}
\newcommand{\sheprev}[1]{{\color{black}#1}}

\newcommand{\gs}[1]{%
  \ifthenelse{\boolean{showcomments}}{%
    \textcolor{BrickRed}{\small[Gururaj: #1]}%
  }{}%
}
\newcommand{\alaacomment}[1]{%
  \ifthenelse{\boolean{showcomments}}{%
    \textcolor{BrickRed}{\small[Comments from Alaa: #1]}%
  }{}%
}
\newcommand{\brian}[1]{%
  \ifthenelse{\boolean{showcomments}}{%
    \textcolor{cyan}{\small[Brian: #1]}%
  }{}%
}
\newcommand{\ms}[1]{%
  \ifthenelse{\boolean{showcomments}}{%
    \textcolor{red}{\small[Mark: #1]}%
  }{}%
}
\newcommand{\oo}[1]{%
  \ifthenelse{\boolean{showcomments}}{%
    \textcolor{red}{\Small[Oleksii: #1]}%
  }{}%
}
\newcommand{\mg}[1]{%
  \ifthenelse{\boolean{showcomments}}{%
    \textcolor{BurntOrange}{[MG: #1]}%
  }{}%
}

\newcommand{\TODO}[1]{%
  \ifthenelse{\boolean{showcomments}}{%
    \textcolor{red}{[TODO: #1]}%
   }{}%
}

\crefname{section}{\S}{\S\S}
\crefname{subsection}{\S}{\S\S}

\copyrightyear{2025}
\acmYear{2025}
\setcopyright{acmlicensed}
\acmConference[ASPLOS '25]{Proceedings of the 30th ACM International Conference on Architectural Support for Programming Languages and Operating Systems, Volume 2}{March 30-April 3, 2025}{Rotterdam, Netherlands}
\acmBooktitle{Proceedings of the 30th ACM International Conference on Architectural Support for Programming Languages and Operating Systems, Volume 2 (ASPLOS '25), March 30-April 3, 2025, Rotterdam, Netherlands}
\acmDOI{10.1145/3676641.3716247}
\acmISBN{979-8-4007-1079-7/25/03}
\ccsdesc[500]{Security and privacy~Security in hardware}
\keywords{Side Channels, Spectre, Defenses, Fuzzing}

\begin{document}

\title{\Gemfuzzer{}: Automated Design-Time Testing of Secure Speculation Countermeasures} 

\author{Bo Fu}
\affiliation{%
  \institution{University of Toronto}
  \city{Toronto}
  \country{Canada}
}
\email{fubof@cs.toronto.edu} 
\authornote{Bo and Leo contributed equally as lead authors for this paper}
\author{Leo Tenenbaum}
\affiliation{%
  \institution{University of Toronto}
  \city{Toronto}
  \country{Canada}
}
\email{leo.tenenbaum@mail.utoronto.ca} 
\authornotemark[1]
\author{David Adler}
\affiliation{%
  \institution{University of Toronto}
  \city{Toronto}
  \country{Canada}
}
\email{d.adler@mail.utoronto.ca} 
\author{Assaf Klein}
\affiliation{%
  \institution{Technion - Israel Institute of Technology}
  \city{Haifa}
  \country{Israel}
}
\email{assafklein@campus.technion.ac.il} 
\author{Arpit Gogia}
\affiliation{%
  \institution{IMDEA Software Institute}
  \city{Madrid}
 \country{Spain}
}
\email{arpitgogia@proton.me} 
\author{Alaa R. Alameldeen}
\affiliation{%
  \institution{Simon Fraser University}
  \city{Burnaby}
  \country{Canada}
}
\email{alaa@sfu.ca} 
\author{Marco Guarnieri}
\affiliation{%
  \institution{IMDEA Software Institute}
  \city{Madrid}
  \country{Spain}
}
\email{marco.guarnieri@imdea.org} 
\author{Mark Silberstein}
\affiliation{%
  \institution{Technion - Israel Institute of Technology}
  \city{Haifa}
  \country{Israel}
}
\email{mark@ee.technion.ac.il} 
\author{Oleksii Oleksenko}
\affiliation{%
  \institution{Azure Research, Microsoft}
  \city{Cambridge}
 \country{United Kingdom}
}
\email{oleksii.oleksenko@microsoft.com} 

\author{Gururaj Saileshwar}
\affiliation{%
  \institution{University of Toronto}
  \city{Toronto}
 \country{Canada}
}
\email{gururaj@cs.toronto.edu} 

\begin{abstract}
In recent years, several hardware-based countermeasures proposed to mitigate Spectre attacks have been shown to be insecure.
To 
\rev{enable the development of} effective secure speculation countermeasures, we need \emph{easy-to-use} tools that can automatically test their security guarantees \emph{early-on} in the design phase to facilitate rapid prototyping.
%

This paper 
develops \Gemfuzzer{}, the first tool capable of testing secure speculation countermeasures for speculative leakage early in their design phase in simulators.
Our key idea is to leverage model-based relational testing tools that can detect speculative leaks in commercial CPUs, and apply them to micro-architectural simulators to test secure speculation defenses.
We identify and overcome several challenges, including designing an expressive yet realistic attacker observer model in a simulator, overcoming the slow simulation speed,
and searching the vast micro-architectural state space for potential vulnerabilities. \Gemfuzzer{}  speeds up test throughput by more than 10$\times$ compared to a naive design and uses techniques to amplify vulnerabilities to uncover them within a limited test budget. 
Using \Gemfuzzer{}, we launch for the first time, a systematic, large-scale testing campaign of \rev{four} secure speculation countermeasures \rev{from 2018 to 2024}---InvisiSpec, CleanupSpec, STT\rev{, and SpecLFB}---and uncover 3 known and \rev{6} unknown bugs and vulnerabilities, within 3 hours of testing.
We also show for the first time that the \sheprev{open-source implementation} of SpecLFB is insecure.

\end{abstract}

\renewcommand{\shortauthors}{Bo Fu et al.}

\maketitle 

\vspace{-0.1in}
\section{Introduction}
Spectre attacks~\cite{Spectre} exploit speculative execution to access architecturally unreachable information and leak them via \uarch{} side channels.
%
To mitigate these leaks, many hardware countermeasures have been proposed using techniques like
{\em invisible speculation}~\cite{invisispec, safespec, delay-on-miss}, \textit{undo-based approaches}~\cite{cleanupspec}, and  {\em tracking speculative flows}~\cite{STT, SDO, DOLMA, NDA}. 
%
Many of these countermeasures, however,  have been shown to be insecure, often in months.
For instance, invisible speculation is vulnerable to \textit{speculative interference attacks}~\cite{SpecInterferenceAttack}, whereas CleanupSpec~\cite{cleanupspec}  and  STT~\cite{STT} were shown to be insecure in later works~\cite{unXpec,DOLMA}.
%
Even defenses proposed by CPU manufacturers have been broken by subsequent attacks 

To enable effective secure speculation defenses, we need tools that can automatically test their security.
These tools need to satisfy two requirements to be practically  adopted:

\begin{enumerate}[left=10pt]
    \item[\textbf{R1:}] They need to  be applicable \emph{early} in the design phase to facilitate rapid prototyping of countermeasures. 
    \item[\textbf{R2:}] They need to be \emph{easy to use}  on existing design artifacts to increase adoption by computer architects.
\end{enumerate}



Currently, computer architects lack tools that meet both requirements.
%
\emph{Formal methods}~\cite{ccs2023_zilong,upec,checkmate,pensieve} for reasoning about \uarch{} leaks have limited scalability and require expertise
which computer architects often lack 
(failing \textbf{R2}).
For instance, Pensieve~\cite{pensieve}
requires formalizing \uarch{} countermeasures 
in a dedicated modeling language.
%
\emph{RTL testing tools}~\cite{borkar2024whisperfuzz,specdoctor}, while scalable and enabling pre-silicon testing, fail \textbf{R1}, as they are inapplicable early in design, when architects prototype features on simulators.
%
A majority of secure speculation countermeasures are \rev{prototyped} in simulators, making RTL-based tools unsuitable for testing them.
In this work, we address this gap with \Gemfuzzer{}, the first tool that can test secure speculation countermeasures at design time in \uarch{} simulators.
\Gemfuzzer{} enables \ul{A}utomated \ul{$\mu$}-architectural \ul{Le}akage \ul{T}esting, \textit{i.e.}, the testing of a countermeasure for unexpected speculative leaks to discover vulnerabilities 
at design time. 
%
For practical adoption, we seek to avoid intrusive changes to either the simulator or the countermeasure being tested. 
%
For this, \Gemfuzzer{} adapts \emph{model-based relational testing (MRT) techniques}~\cite{revizor,HideSeekSpectre},
which found speculative leaks in commercial CPUs, to 
\uarch{} simulators.

Following the MRT approach, \Gemfuzzer{} tests the target defense (implemented on a \uarch{} simulator) against a given \textit{leakage contract}~\cite{hw_sw_contracts}, an ISA-level model capturing the expected leakage.
\Gemfuzzer{} has two main components:  (a) a \textit{model} that maps program executions to \textit{contract traces}, i.e., sequences of ISA-level observations capturing the expected leakage according to the contract, and (b) an \emph{executor} that generates \textit{\utrace{}s}, which capture the observable side-effects of speculative execution on the simulator.
\Gemfuzzer{} generates random programs executed on both the executor and the model, and compares the expected leaked information (captured by the contract traces) with the actual information leaked by the defense (captured by the \utrace{}s).
Any discrepancy between the two (called a \emph{contract violation}) indicates an unexpected leak, and therefore a potential security vulnerability in the defense.





%
While we reuse the test and contract trace generation from prior work~\cite{revizor,HideSeekSpectre}, implementing  \tool{} requires addressing three core challenges (indicated as C1--C3 below).


\vspace{0.05in}
{\bf \noindent C1: What CPU state to expose in the \utrace?} 
The design of the \utrace{} is critical in MRT as it captures the observational power of the attacker against which the security guarantees are tested.
Simulators allow the definition of extremely 
expressive \utrace{}s comprising potentially the entire CPU \uarch{} state.
At the same time, \utrace{} needs to be grounded in a realistic observer model ,
so that the detected violations may be exploitable on a real CPU. 

In \tool{}, we implement \utrace{}s by taking a snapshot of the final cache and TLB states of the test program extracted from the simulator. 
%
%
Our evaluations \rev{in \cref{sec:eval-def}} show this \utrace{}, capturing the observational power of an attacker exploiting memory-system side channels \rev{(\textit{explicit channels}~\cite{STT})}, is sufficient to discover exploitable speculative leaks in several secure speculation countermeasures \rev{claiming to protect against these side channels}. At the same time, it is also easily extensible to other attacker models. \rev{In \cref{sec:trace_formats}, we show that exposing more information in the \utrace{s}, e.g., branch predictor state or 
program counter sequence, can also detect \textit{implicit channels} based on branch prediction or resolution,  at the cost of reduced testing throughput.}



\vspace{0.05in}
{\bf \noindent C2: Slow testing speed}. Running test cases in a simulator is slow. This severely limits the number of tests that can be executed during a campaign and reduces the chances of finding contract violations. Our key observation is that, counterintuitively, the main bottleneck is the \emph{startup costs} of the simulator rather than the test runtime. 
This is because current MRT techniques~\cite{revizor,HideSeekSpectre,scamv} generate small test programs (a few tens of instructions) that run quickly (e.g., tens of milliseconds in gem5 \cite{gem5}), whereas the simulator startup times are nearly two orders of magnitude higher.

To address this, we design a testing harness that runs successive tests without restarting the simulator by overwriting register and memory values between the tests, thereby amortizing the startup costs across multiple tests. Compared to restarting the simulator on each test, \tool{} improves test throughput by over 10x, as shown in \Cref{sec:htrace_design}.

\vspace{0.05in}
{\bf \noindent C3: Low probability of discovering leaks}. A key requirement for speculative leaks is contention on \uarch{} resources. To amplify the chances of their occurrence, we test the design with smaller \uarch{} structure sizes (e.g., fewer cache ways, fewer MSHRs), amplifying  contention and making leaks easier to discover.
With this, \tool{} uncovers vulnerabilities that could not be detected otherwise, as shown in \Cref{sec:results_amplify_vio}.\looseness=-1

\noindent
\textbf{Evaluation.}
We use \tool{} to run the first systematic, large-scale testing campaign against secure speculation countermeasures (all implemented in the gem5 simulator).
Our campaign covers an unprotected out-of-order CPU and 
\rev{four} secure speculation countermeasures: InvisiSpec~\cite{invisispec}, CleanupSpec~\cite{cleanupspec}, STT~\cite{STT}\rev{, and SpecLFB~\cite{speclfb}}.
We detect Spectre-v1 and Spectre-v4 vulnerabilities within minutes in the unmodified (insecure) CPU. In the countermeasures under test, we discover  
3 unknown implementation bugs, \rev{3} unknown vulnerabilities in the designs, and confirm 3 known vulnerabilities, all within three hours of testing. 
%

In InvisiSpec, we discover a previously unknown bug in the cache eviction logic, which leaks speculatively accessed addresses via evictions. We also discovered a stronger variant of speculative interference attack~\cite{SpecInterferenceAttack}.  
While the prior attack requires a multi-threaded attacker and SMT support, our variant is exploitable by a single-threaded adversary, thereby breaking InvisiSpec's \rev{security} in a single-threaded setting. 


In CleanupSpec, we discovered several new bugs and vulnerabilities. We find leaks of speculatively accessed addresses due to (a) incorrect cleanups of non-speculative load addresses when they match with reordered speculative loads, a previously unknown vulnerability,
and (b) a lack of cleanup for speculative stores and speculative requests \rev{crossing} cache lines, due to implementation bugs.
We also re-discovered unXpec~\cite{unXpec}, a previously known vulnerability. 


In STT, \Gemfuzzer{} automatically flagged a known vulnerability on speculative stores where the TLB is speculatively accessed by tainted stores, as shown previously by DOLMA~\cite{DOLMA}. 

\rev{
In SpecLFB, a defense proposed in 2024, \tool{} discovers a new vulnerability, similar to Spectre-v1, \sheprev{in the open-sourced gem5 implementation}. This is due to an undocumented optimization in the implementation that removes protection for the first speculative load in the load-store queue, thereby making it vulnerable to Spectre attacks leaking secret registers with a single speculative load.}

\smallskip
\noindent
\textbf{Summary of contributions:} 
\begin{enumerate}

\item  We introduce \Gemfuzzer{}, the first tool
that can automatically find information leaks in simulated CPUs, and identify weaknesses in the proposed designs of secure speculation countermeasures. 


\item We expose the \uarch{} state realistically observable by an attacker, and introduce techniques to amplify the observability of violations in white-box simulators without intrusive changes to the simulator. 


\item We identify performance bottlenecks and address them by streamlining the test case execution, improving the testing throughput by an order of magnitude.

\item We launch the first large-scale testing campaigns on several recent secure speculation countermeasures, InvisiSpec, CleanupSpec, STT\rev{, and SpecLFB,} and automatically find multiple unknown bugs and vulnerabilities in \rev{three} hours of testing on a commodity server.
\end{enumerate}

We have \rev{responsibly disclosed} our discoveries to the authors of the countermeasures.
\tool{} is open-sourced at \url{https://github.com/sith-lab/amulet} to aid the testing of future countermeasures as they are designed.

\section{Background: Testing for speculative leaks}\label{sect:background}

\newcommand{\htrace}{\utrace}
\newcommand{\htraces}{\htrace{}s}
\newcommand{\contract}{\mathit{C}}
\newcommand{\HTrace}{\mu\mathit{Trace}}
\newcommand{\ctrace}{\mathit{CTrace}}

%
We first discuss leakage contracts~\cite{hw_sw_contracts}, which model speculative leaks at the ISA level (\Cref{sect:background:contracts}),
then describe how attackers can be modeled (\Cref{sect:background:uarch-traces}),
%
and how leakage contracts enable detecting unexpected leaks visible to attackers (\Cref{sect:background:contract-violations}).
Finally, we overview Revizor~\cite{revizor} (\Cref{sect:background:revizor}), a testing tool for detecting leaks in CPUs, which we use as a basis for \tool{}.

\subsection{Leakage contracts modeling speculative leaks}\label{sect:background:contracts}


\emph{Leakage contracts}~\cite{hw_sw_contracts} capture the expected \uarch{} leaks at the ISA level.
A contract $\contract$ describes, for any program $p$ and input $i$,  what information might be leaked microarchitecturally when executing $p$ with input $i$.
For this, contract $\contract$ maps each execution to a \emph{contract trace}, i.e., a sequence of ISA-level observations capturing leaked information.
$\contract(p,i)$ denotes the contract trace for the execution of program $p$ with input $i$.\looseness=-1

Contracts are formalized by annotating ISA instructions with (a) an \emph{observation clause} modeling the information leaked by the instruction, and (b) an \emph{execution clause} modeling if (and how) instructions trigger speculation.
\Cref{table:contracts} summarizes the contracts we use in evaluations in \Cref{sec:eval}:

\begin{asparaitem}
\item The \texttt{CT-SEQ} contract models the leakage expected by a CPU with cache side channels and without speculative execution.
Its observation clause exposes the addresses of executed load and store instructions as well as the program counter throughout the execution.
The execution clause is empty, indicating that no instruction triggers speculation and that the leakage is only on {architectural} execution paths.

\item The \texttt{CT-COND} contract models the leakage expected by a CPU with branch prediction.
While its observation clause is the same as for \texttt{CT-SEQ}, the execution clause specifies that when a conditional branch is executed, the corresponding mis-predicted branch should be explored as well.
This captures 
instructions transiently executed due to branch prediction.

\item Finally, the \texttt{ARCH-SEQ} contract exposes the program counter, the location of all loads and stores, and the values of all data loaded from memory on architectural program paths.
For this, \texttt{ARCH-SEQ} extends the observation clause in \texttt{CT-SEQ} by additionally exposing the values loaded from memory.
As in \texttt{CT-SEQ}, the execution clause is empty.
\end{asparaitem}


\begin{table}[htb]
\centering
\vspace{-0.05in}
\caption{Leakage contracts used in this work. }
\vspace{-10pt}
\footnotesize
\begin{tabular}{|m{1cm}|m{3cm}|m{2.7cm}|}
\hline
\multirow{2}{*}{\textbf{Name}} & \multicolumn{2}{c|}{\textbf{Clauses}} \\ \cline{2-3}
&  \textbf{Leakage} & \textbf{Execution}  \\ 
\hline
\texttt{CT-SEQ} & \texttt{PC}, \texttt{LD}/\texttt{ST} \texttt{ADDR} & N/A  \\
\hline
\texttt{CT-COND} & \texttt{PC}, \texttt{LD}/\texttt{ST} \texttt{ADDR} & Mispredicted Branches  \\
\hline
\texttt{ARCH-SEQ} & \texttt{PC}, \texttt{LD}/\texttt{ST} \texttt{ADDR} and values & N/A  \\
\hline
\end{tabular}

\vspace{-0.1in}
\label{table:contracts}
\end{table}

\subsection{Modeling attackers in speculation-based attacks}\label{sect:background:uarch-traces}

In speculation-based attacks, attackers extract information about a victim program by observing changes in the CPU's \uarch{} state through side channels~\cite{PrimeProbe,yaromFlushReload}.
%
Like prior works~\cite{hw_sw_contracts,revizor}, we model attacker observations using \htraces{} from victim execution, where each trace captures the \uarch{} changes observed by an attacker. 
%
%
%
For a program $p$ starting from an input $i$ and \uarch{} context $\mu$ (the initial CPU \uarch{} state), $\HTrace(p,i,\mu)$ denotes the \htrace{} for this execution.
%
In~\Cref{sect:design}, we describe the \htraces{} used in \tool{}. 

\subsection{Detecting unexpected leakages}\label{sect:background:contract-violations}

A leakage contract captures the \emph{expected leakage} of a CPU under test.
Any \emph{unexpected} leak is a contract violation, where the \utrace{}s leak more information than the contract traces. 
This is defined precisely as follows: 


\vspace{-0.025in}
\begin{definition}[Contract violation~\cite{hw_sw_contracts}]\label{def:violation}
A CPU violates a contract $\contract$ if there exists a program $p$, two inputs $i,i'$, and a microarchitectural context $\mu$ such that $\contract(p,i) = \contract(p,i')$ and $\HTrace(p,i,\mu) \neq \HTrace(p,i',\mu)$.
\end{definition}
\vspace{-0.025in}

%
%
A violation is evidence of an unexpected leak \rev{in the CPU} as the attacker can distinguish two executions ($\HTrace(p,i,\mu) \neq \HTrace(p,i',\mu)$) that should be indistinguishable based on the contract ($\contract(p,i) = \contract(p,i')$).

\subsection{Model-based relational testing: Revizor}\label{sect:background:revizor}

Model-based relational testing (MRT) tools
~\cite{revizor,scamv}
discover unexpected leaks in CPUs by searching for contract violations,
\rev{and have been successful in finding} new vulnerabilities and variants of 
\rev{existing} 
ones~\cite{revizor,HideSeekSpectre,hofmann2023SpecAtFault,scamv}.
Here, we focus on Revizor, the MRT tool we use as a basis for \tool{}.

Revizor searches for  contract violations by (1) generating a random program $p$ and a sequence of random inputs $[i_0, i_1, \ldots]$, (2) collecting the contract and \htraces{} for all programs and inputs, and (3) analyzing the traces to identify violations according to \Cref{def:violation}.
The testing continues for a fixed number of rounds or until a violation is detected.
We now provide further details on the core parts of Revizor:
\begin{asparaitem}
    \item \emph{Program generation}: 
    The generator selects a random sequence of assembly instructions from a pool to form a program.
    It can be configured to constrain the shape of the program's control-flow graph, control the pool of instructions, and configure the instruction frequencies.
    
    \item \emph{Input generation}:
    Each input is a binary file, generated with a (seeded) pseudo-random number generator, that initializes the test program's memory and registers.
    %
    Inputs can also be mutated, 
    preserving only the parts influencing the contract trace while randomizing others to ensure identical contract traces but potentially different speculative behavior.
    
    \item \emph{Collecting contract traces:} 
    Revizor implements an executable version of the contract (called \emph{leakage model}) on top of the Unicorn ISA emulator~\cite{Unicorn} by (1) adding instrumentation to record observations according to the contract's observation clause, and (2) simulating speculative execution paths as per the execution clause.
    Revizor collects contract traces by executing the program $p$ with all the inputs  $[i_0, i_1, \ldots]$ using the leakage model and recording the observations.

    \item \emph{Collecting \htraces{}:}
    Revizor implements an \emph{executor} that takes a program $p$, executes it on the target CPU with each of the inputs $[i_0, i_1, \ldots]$, and measures the \uarch{} trace for each execution. 
    These traces are collected via a side channel attack, like Prime+Probe, where each trace is a set of cache lines evicted by the program.
    %

    \item \emph{Comparing leakage}:
    At the end of each testing round, Revizor compares the collected contract and \htraces{} to detect violations according to \Cref{def:violation}.
\end{asparaitem}

\section{\Gemfuzzer{} Design}\label{sect:design}

\Gemfuzzer{} builds on Revizor~\cite{revizor}, an MRT tool that finds leaks in silicon CPUs, \rev{to enable}
design-time testing of secure speculation mechanisms \rev{in} simulators.
Next, we provide a high-level overview of \tool{}, highlight the key challenges in testing designs in a simulator, and explain our solutions. 

\subsection{Overview of \tool{}}

\begin{figure}[tb]
    \includegraphics[width=3.3in]{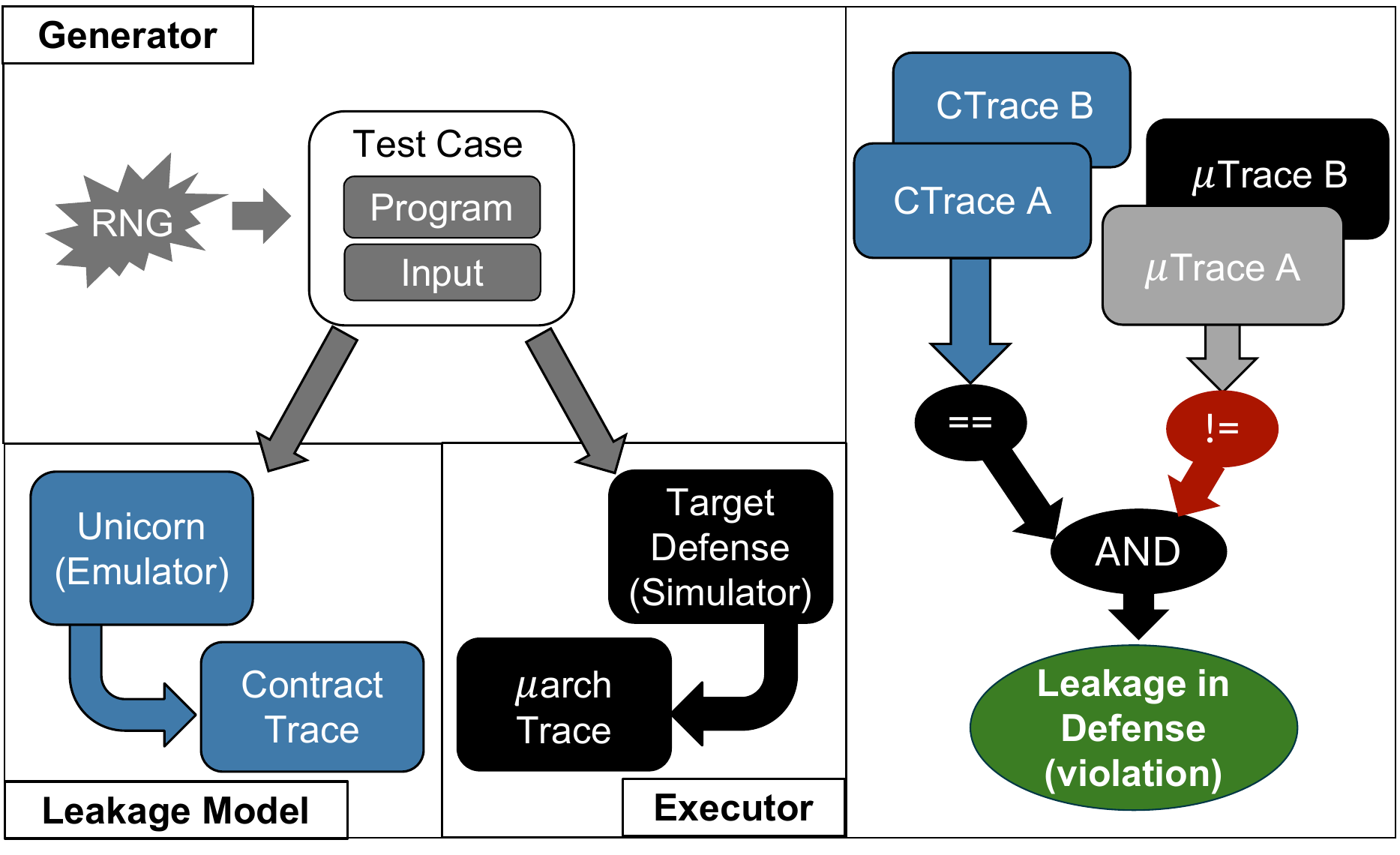}
    \caption{Overview of \tool{}. We leverage the test-generation and leakage models from prior works~\cite{revizor,HideSeekSpectre,hofmann2023SpecAtFault} and design a new executor in \tool{} capable of testing target defenses in a \uarch{} simulator.}
    \label{fig:overview}
\end{figure}


\tool{} consists of three main modules, as shown in \cref{fig:overview}: (1) the \textit{test generator}, which generates random programs and inputs, (2) the \textit{leakage model}, which generates the contract traces, 
and (3) the \textit{executor}, which generates \utrace{}s from a simulator implementing the countermeasure under test.
Below, we describe each component briefly.

\paragraph{Test Generator.}
\tool{} reuses the test generator from Revizor~\cite{revizor}  (described in \cref{sect:background:revizor})
 to generate short test programs of up to 5 basic blocks of randomly selected instructions, linked together by jumps in the form of a directed acyclic control flow graph. All  memory accesses are forced to access a predefined and initialized memory sandbox. We vary the number of 4KB pages in the sandbox from 1 to 128. 
Several random inputs are generated for each test program.  A combination of a program and input forms a test case.\looseness=-1

\paragraph{Leakage Model}
\tool{} reuses the leakage model from Revizor~\cite{revizor} to collect contract traces by executing each test case  on the Unicorn~\cite{Unicorn} CPU emulator.
For each of the target defenses, we test them against a contract that matches their security guarantees, following the formal analysis by Guarnieri et al.~\cite[Section VI]{hw_sw_contracts}. Specifically, 
we use the \texttt{CT-SEQ} contract for testing InvisiSpec~\cite{invisispec} (Futuristic), CleanupSpec~\cite{cleanupspec}\rev{, and SpecLFB~\cite{speclfb},} and the \texttt{ARCH-SEQ} contract for testing STT~\cite{STT} (Futuristic).

\paragraph{Executor}
The executor generates the \utrace{}s from a simulator implementing the countermeasure under test.
\utrace{}s model attacker observations for a given countermeasure.
When two tests have matching contract traces but different \utrace{}s, we flag that as a contract violation (cf.~\Cref{def:violation}), indicating a speculative leak (and a potential vulnerability).
Thus, the design of the \utrace{} is critical in determining the types of leaks that can be detected. 

\subsection{\uarch{} Trace Design - Challenges and Solutions}\label{sec:htrace_design}
As the \utrace{}s model the attacker observations, they play a critical role in determining the speed and efficacy of testing for leaks. Below, we highlight the key challenges in designing the \utrace{} in \tool{} and then our solutions.

\vspace{0.05in}

\noindent \textbf{C1. Determining $\mu$Arch States Exposed in $\mu$Arch Trace}

\noindent A simulator provides white-box access to the \textit{entire} CPU's \rev{\uarch{}} state, which can be potentially exposed via the \utrace{}. 
\rev{However, exposing information that is too detailed may reduce the testing throughput.
Moreover, although} more detailed \utrace{}s can result in more \rev{contract} violations \rev{(cf.~\cref{def:violation})}, not all discovered violations may \rev{lead to} exploitable leaks for \rev{software-based} attackers.
%

Consider \rev{the following} options for \rev{the} \utrace{}, \rev{which expose different \uarch{} information.} 
The first uses a snapshot of the cache and TLB state at each test case's end (i.e., L1D-cache and D-TLB tags).
This models a realistic software-based attacker inferring the cache or TLB state \rev{by performing memory accesses and checking for cache or TLB hits/misses (an \textit{explicit channel} as per the STT taxonomy~\cite{STT})}.
\rev{
The second option uses a snapshot of the branch-predictor (BP) state (consisting of local/global history tables, branch target buffer, etc.) at the end of the test case. 
This models a more sophisticated attacker 
that can infer information about \textit{how} program   branches have been executed from the secret-dependent BP state (this is an example of an \textit{implicit channel based on prediction} in STT's taxonomy).
}

\rev{Finally,} the \rev{third} option models an attacker \rev{physically probing the hardware, i.e., monitoring  \uarch{} state transitions throughout the execution of the test case.
For example, monitoring the sequence of  program counter values (PC) and branches, or monitoring each transaction on the L1-Cache bus exposing the sequence of addresses of each load/store}. This provides much more precise and exhaustive information.
\rev{For instance, differences in the sequence of PCs can detect \textit{implicit channels} based on branch resolution (as per STT's taxonomy), whereas a secret-dependent reordering of loads may induce a difference in the cache replacement state.
However, not all of these transient differences may be exploitable in practice; some of these secret-dependent orderings may not result in changes observable by a realistic attacker 
(e.g., reordered loads map to different cache sets).

In \cref{sec:trace_formats}, we evaluate testing campaigns using each of these three \utrace{} formats and show that \tool{} can effectively detect different kinds of violations regardless of the trace format.
}
%
%
%
\rev{
By default, in \tool{}, we use state-based \utrace{s} to model the most realistic attacker.
We use the data cache and TLB snapshots (first option) for the 
\utrace{}s, given that such memory-system-based side channels are exploited 
in the majority of speculation-based attacks and most defenses~\cite{cleanupspec,invisispec,speclfb,STT} that we seek to test with \tool{} protect against at least these side channels.}
%

\begin{tcolorbox}[left=1mm, right=1mm, top=1mm, bottom=1mm, boxsep=1mm]
In \tool{}, we use the \utrace{} comprising a snapshot of the final TLB and caches state, \rev{which} models a realistic attacker observing memory-system side channels.  
\end{tcolorbox}

Our evaluation \rev{in \cref{sec:eval-def}} shows that such a \utrace{} provides sufficient information to discover exploitable violations \rev{across a wide variety of defenses.}
\rev{We remark, however, that any other \uarch{} state observable to an attacker (e.g., second or third option) can also be used as part of the \utrace{}.}

\ignore{
While this does not leverage extra \uarch{} information available in the simulator,
we show how we can compensate for that and amplify the observability of leakages using the white-box nature of the simulator, in \cref{sec:amplify_vuln}. 
}


\ignore{
This gives us all the physical memory addresses of lines in the cache when the test case exits, but does not tell us which instructions brought any given line into cache. 
We define a contract violation to occur here for a pair of test cases when the contract traces match, but the final cache state of any cache (L1D, L1I, L2) does not. We consider this stronger evidence of incorrect speculation, as it is comparable to what will be seen by a Prime+Probe attacker model.  This is known as the "state trace". 

Second, we can tap into the channel where memory accesses are sent into the cache hierarchy. This allows us to see which individual instruction brought a line into cache, the PC of that instruction (s.t. we can discern between multiple instructions with the same PC), and the physical memory address of the new cache line. We can also observe the tick at which that access occurred, whether that access was a read or write, and/or the size of data requested by that access. 

This approach differs from the state trace as it allows us to see differences in memory access ordering. Here, a contract violation is defined to occur when for a test case pair, we observe a different memory access ordering. However, we consider this weaker evidence of incorrect speculation, as it may reveal differences in timing that are impractical or impossible to observe. A realistic attacker is most likely unable to snoop directly into the cache network the way we are able to do so on gem5. This is known as the "ordered trace"

We can augment the ordered trace with the timing of when specific accesses occur, although this is not done by default. In this case, a contract violation is defined even more strictly, as a difference in timing of when memory accesses are sent out. This is much harder for a real observer to discern as there may be no differences in cache access order, and some differences in memory timing can be as small as a few 10s of cycles. We call this the "timing trace"

In the interest of realistic observability, we used only the state trace to find violations for the results we will discuss here. All of the violations we detail should be easily observable by a Prime+Probe based attacker. 
}

\vspace{0.05in}
\noindent \textbf{C2. Determining Initial State for $\mu$Arch Traces.}
Prior works~\cite{revizor,HideSeekSpectre} collect \utrace{}s on real CPUs using cache side channels like Flush+Reload~\cite{yaromFlushReload} or Prime+Probe~\cite{PrimeProbe}. 
While we directly extract the final state of the cache from the simulator and do not need to infer it via a side channel, this still leaves the question: What should be the initial state of the caches before starting a test case? 

The most intuitive approach is to start deterministically from a clean cache state, thus eliminating noise.

\begin{tcolorbox}[left=1mm, right=1mm, top=1mm, bottom=1mm, boxsep=1mm]
We observe that the best results are obtained when initializing the L1 cache by filling it up with addresses from outside the memory sandbox of the test case. This translates into  64 x 8 addresses for an 8-way, 32KB L1 cache.
\end{tcolorbox}

Starting from a fully occupied cache set ensures we not only detect leakages due to speculative cache line installs (addresses installed by the program), but also due to replacements (\rev{evicted addresses}). 
In our evaluation (\cref{sec:naive_vs_opt}), we show that initializing the cache state in this way increases the number of detected violations compared to the naive approach.


\vspace{0.05in}
\noindent \textbf{C3. Alleviating Slowdown Due to Slow Simulator Startup.}

\noindent While MRT tools for testing silicon CPUs face a performance bottleneck due to test and contract trace generation~\cite{HideSeekSpectre}, the bottleneck for testing countermeasures in simulators is the simulation runtime, i.e., the time to execute the test and extract the \utrace{}. 
The naive approach of generating a \utrace{} for each test program and input involves packaging the test case (program and input) in a binary, running the binary on the simulator, and extracting the \utrace{} at the end of the simulation.
Unfortunately, we observe that this approach is not very performant due to the use of short test cases (approximately 50 instructions per test case)\footnote{Short test cases are beneficial to ensure that side effects of speculation are observable in the final cache state obtained at the end of the test case.}.



\smallskip
\textbf{\textit{Characterizing \Gemfuzzer{}-Naive.}}
We analyze the execution time for a single test program in \cref{table:breakdown}. We run tests on the default Out-of-Order CPU (non-secure) in gem5, which we run in SE mode. We see that 97\% of the time is spent in gem5 as expected; however, 96.1\% of the time is spent in gem5 start-up, and only 0.9\% is spent simulating the instructions. This is because gem5 takes a few seconds to initialize,
but just tens of milliseconds to run a test case of \char`\~50 instructions.
Thus, the startup time is the primary bottleneck.

\begin{table}[bht]
\centering
\small
\caption{
Breakdown of time per test program, using a Naive and Optimized \utrace{} extraction in \Gemfuzzer{} (with 140 inputs/program), in a campaign running 30 test programs.}
\begin{tabular}{lcc}
\hline 
\textbf{Component}&\textbf{Naive} &\textbf{Opt}\\ \hline 
gem5 startup    & 156 s (96.1\%)   & 0.2 s (1.6\%)\\
gem5 simulate   & 1.4 s (0.9\%)      & 11 s (88.5\%)\\
$\mu$Trace extraction & 0.9 s (0.5\%)    & 0.6 s (4.6\%)\\
Test generation   & 0.5 s (0.3\%)    & 0.3 s (2.5\%)\\
CTrace extraction & 0.1 s (0.1\%)     & 0.1 s (0.6\%)\\
Others            & 3.4 s (2.1\%)   & 0.3 s (2.2\%)\\\hline
\textbf{Total}    & 159 s (100\%)  & 12 s (100\%)\\\hline
\end{tabular}
\label{table:breakdown}
\end{table}

\begin{figure}
    \includegraphics[width=3.3in]{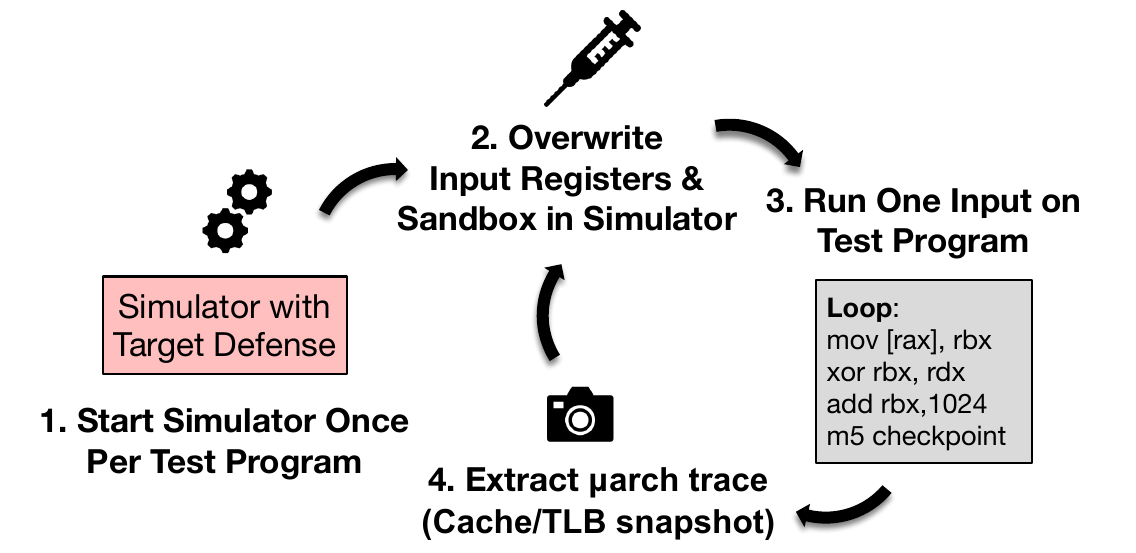}
    \caption{Design of \utrace extraction in \tool{}-Opt.}
    \label{fig:amulet_opt_design}
\end{figure}

\textbf{\textit{\Gemfuzzer{}-Opt.}} We \rev{adopt} an optimized \utrace{} extraction method in \Gemfuzzer{} (Opt in \cref{table:breakdown}).
%
%
Rather than restarting the simulation for each program input, we \rev{continue to} execute test cases for different inputs of the same program 
\rev{by directly overwriting} the register \rev{and memory} values in the simulated binary in the gem5 process 
without restarting the simulator. 
As shown in \cref{fig:amulet_opt_design}, we start the simulation with a binary that loops over the test program instructions. After each successive input, we overwrite the input register values in the binary and continue the simulation.
At the end of each iteration, we extract the corresponding \utrace{}.

\begin{tcolorbox}[left=1mm, right=1mm, top=1mm, bottom=1mm, boxsep=1mm]
Unlike \Gemfuzzer{}-Naive, which restarts the simulator for each input, \Gemfuzzer{}-Opt only
restarts it for each test program,
so the startup cost 
\rev{is} amortized across all inputs.
\end{tcolorbox}

\textit{\textbf{Benefits of \tool{}-Opt.}} As shown in the  \textit{Opt} column of \cref{table:breakdown},  the gem5 startup time is significantly reduced (consuming 2\% of the time), and the bottleneck is now the time for simulating instructions (consuming 89\% of the time).
Note that our simulation time per program increases, since we need 10x more instructions to reset the cache state per test input with addresses from outside the sandbox (64 $\times$ 8 instructions for a 8-way, 32KB L1D-cache). We considered adding a special custom instruction to reset the cache to further reduce the time per test but ruled out this idea to avoid intrusive changes to the design that may potentially change its behavior.
The time per test program for \Gemfuzzer{}-Opt is just 12 seconds, 13x lower compared to \Gemfuzzer{}-Naive, which takes  2.7 minutes per test program with all its inputs.  

\tool{}-Opt has an additional benefit that it preserves
the \uarch{} state of the predictors (like branch predictor or memory dependence predictor) 
between test cases. This is beneficial for
finding leaks \rev{as} it results in a wider variety of predictions in successive inputs. 
However, this also means that a violation might be \rev{due to} differences in the initial \uarch{} context. 
Following prior work~\cite{revizor}, we \textit{validate} a violation by 
re-running the violating inputs with the other test case's \uarch{} starting context and check if the violation persists.
%
%
We provide detailed comparisons of the speed of testing and efficacy in finding violations between Naive and Opt in \cref{sec:naive_vs_opt}.



\subsection{Analyzing Violations}\label{sec:analyzing-violations}
\begin{figure}[tb]
    \includegraphics[width=3.0in]{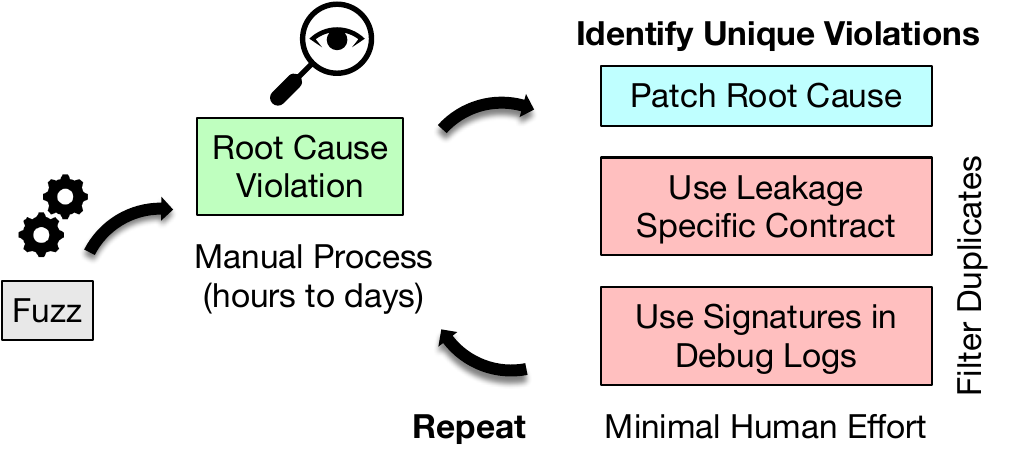}
    \caption{\sheprev{Analyzing violations discovered with \tool{}.}}
    \label{fig:amulet_vio_analysis}
\end{figure}

Violations \rev{are} detected by \Gemfuzzer{} \rev{when there is} a difference in the \utrace{}s (final \rev{D-}cache or TLB state) for two inputs to a program \rev{with} the same contract trace (cf. \Cref{def:violation}). 
\rev{On detecting a violation, \Gemfuzzer{} outputs the program and the pair of inputs causing the violation with their \utrace{}s. 
}
\rev{
Our violation analysis workflow, shown in \cref{fig:amulet_vio_analysis}, has two steps: (a) root cause analysis of a given violation and (b) identifying unique violations by filtering out similar ones.\looseness=-1

\smallskip
\noindent {\bf (a) Root Cause Analysis.}
\sheprev{Like any fuzzing approach, identifying the root cause of a violation in \Gemfuzzer{} is a manual process. This is because fuzzing typically automates the testing process, whereas root causing detected vulnerabilities is orthogonal and often dependent on the vulnerability itself.

We analyze violations in \Gemfuzzer{} that manifest as differences in the D-Cache and TLB-based \utrace{s}, by} first identifying the load instruction responsible for the differing addresses in the \utrace{s}. We then trace back along the program data flow to find the mis-speculated instruction dependent on varying inputs (the source of the leak). This identifies the entire mis-speculated instruction sequence causing the violation. We do this using a script that parses gem5’s debug logs (containing information like load/store addresses, branch prediction, etc.) and provides a side-by-side comparison of memory accesses under the two violating inputs and highlights differences. It also displays squashes, which helps locate the cause of the mis-speculation. Once the instruction sequence causing the violation is thus identified,
we examine the gem5 debug logs for this sequence to pinpoint the code in the defense causing the leakage.
\sheprev{For violations discovered in this work, root cause analysis took a few hours to few days, based on the complexity of a violation.}

}

\vspace{0.05in}
\smallskip
\noindent {\bf (b) Identifying Unique Violations.} 
Once we root cause a given violation, we avoid re-discovering similar violations in two ways: (1) fixing the root cause when possible, (2) filtering similar violations using \sheprev{contracts that expose this leakage, or by inspecting} the debug logs. \sheprev{When fixing the vulnerability in gem5 took less than 10 lines of code, we wrote a patch and re-ran the violations to see which got resolved---this worked for violations UV1 and UV3 in InvisiSpec and CleanupSpec (\cref{sec:IS-violations,sec:CS-violations}) with minimal effort}. 

If the vulnerability was fundamental to a defense and hard to patch, \sheprev{we used a contract exposing this leakage in the contract trace to filter these violations, as in prior works~\cite{revizor,HideSeekSpectre}---this approach worked for KV1 and UV6 in InvisiSpec and SpecLFB (\cref{sec:IS-violations,sec:SpecLFB-violations}) and required few lines of code in \tool{}. When characterizing leaks at the contract level was not straightforward, we identified a unique signature for a violation based on patterns in the \utrace{s} or simulator debug logs and used regex-based scripts to filter similar violations. 
For example, InvisiSpec violation (UV2; \cref{sec:IS-violations}) shows MSHR-related stalls in debug logs due to speculative interference, while CleanupSpec violation (UV4; \cref{sec:CS-violations}) shows load requests crossing cache line boundaries in debug logs.}
This approach, which is akin to defining a leakage-specific contract, isolates violations with similar signatures, \sheprev{ and worked for the remaining violations in InvisiSpec and CleanupSpec (\cref{sec:IS-violations,sec:CS-violations})}. We continue steps (a) and (b) until all violations are root caused.
\sheprev{When left with a handful of violations in a defense (e.g., less than 10), we manually inspect the debug logs to verify that all of them have the same signature---this approach worked for  STT   (\cref{sec:stt-violations}).
Overall, identifying unique violations takes minimal manual effort.}

\subsection{Amplifying Leakages in Simulators}\label{sec:amplify_vuln}

So far, \Gemfuzzer{} directly tests any CPU design or simulator-based countermeasure for leaks without any modifications, as if it were a black-box.
However, programs that induce leakages can be fundamentally hard to find with random testing, and exhaustively traversing the vast \uarch{} state space of a design to check for leakage is impractical.

To efficiently uncover such leaks, we leverage the fact that observing speculative leakage has two requirements: (1) a speculative ``access'' instruction that reads the data to be leaked, and a (2) speculative ``transmitter'' instruction, that leaks the data via contention on a \uarch{} resource (i.e., a covert channel) and impacts the \utrace{}. 
%
To make leaks easier to observe, we increase the chance of contention on \uarch{} structures by configuring the target design with smaller \uarch{} structure(s)
(e.g., smaller L1-Caches, fewer cache ways, fewer MSHRs). 
This approach makes covert channels, where contention is unlikely to occur in short test cases (e.g., Prime+Probe, MSHR contention), more likely, thus increasing the chance of detecting speculative leaks.

Note that we do not modify the defense itself or test infeasible CPU configurations.
We use valid configurations of these structures with reduced sizes to increase the probability of short test cases inducing contention. This will make any speculative leaks that exist in the design more observable. 

In evaluations, we show that \tool{} can already discover leaks in countermeasures while testing with default configurations, as we show in \cref{sec:eval-def}; but it can discover more interesting leaks with amplification, as we show in \cref{sec:results_amplify_vio}.


\subsection{Implementation}

We integrate \Gemfuzzer{} with the gem5 simulator~\cite{gem5}.
Our test and contract trace generation build on Revizor's implementation~\cite{revizor}.
We \rev{add} \utrace{} extraction \rev{to} the gem5 \rev{code} of each  defense, and run gem5 in Syscall Emulation (SE) mode.

For the \utrace{}, we use the final state of the L1D-cache and D-TLB extracted from the simulator at the end of a test case.
We reset the L1D-cache state after each test case by filling it with addresses from pages outside the memory sandbox, which also evicts the TLB entries \rev{(for InvisiSpec and STT), or by invalidating the caches directly using a simulator hook if this is supported by the specific simulator version (for SpecLFB and CleanupSpec).
}
For InvisiSpec, CleanupSpec\rev{, and SpecLFB}, we use a sandbox with 1 physical page (as the TLB is not protected), whereas for STT we use a sandbox with 128 pages since we seek to test it also for TLB leaks. 

\section{Evaluation}\label{sec:eval}

\newcommand{\testcase}{test case}
\newcommand{\testcases}{\testcase{}s}

Here, we answer the following research questions:

\begin{rqenumerate}
    \item Can \Gemfuzzer{} detect known leaks in non-secure CPUs? How fast \rev{is} \opt{} compared to \naive{}?
    \item \rev{How does the choice of the \utrace{} format affect the number and kinds of violations found?}
    \item Can \Gemfuzzer{} detect known and unknown leaks in defenses?
    How effective is our leakage amplification? 
    \item Can \Gemfuzzer{} detect more interesting and more fundamental vulnerabilities, as we fix the simpler ones?
    
\end{rqenumerate}

We answer RQ1 \rev{and RQ2} by testing the Out-of-Order gem5 CPU with \Gemfuzzer{} (\S~\ref{sec:naive_vs_opt}--\ref{sec:trace_formats}), and \rev{RQ3} and \rev{RQ4} by testing InvisiSpec, CleanupSpec, STT\rev{, and SpecLFB} (\S~\ref{sec:eval-def}--\ref{sec:stt-violations}). 

\subsection{Evaluation Methodology}

We use \Gemfuzzer{} to test the baseline Out-of-Order CPU in gem5 (O3CPU) and \rev{four} different countermeasures: InvisiSpec~\cite{invisispec}, CleanupSpec~\cite{cleanupspec}, STT~\cite{STT}\rev{, and SpecLFB~\cite{speclfb}}.
For the baseline, we test the \rev{insecure} O3CPU in the gem5 code-base from InvisiSpec.
For each countermeasure, we test its publicly available implementation, run with the default configuration flags.
For InvisiSpec and STT, we run them in their Futuristic mode, as it provides the strongest security.

We run our testing \rev{campaigns} on an AMD EPYC 128-Core CPU.
For comparisons between \tool{}-Naive and -Opt (\cref{sec:naive_vs_opt}), due to the slower speed of \tool{}-Naive, we run shorter test campaign of 16 parallel instances of \Gemfuzzer{}, 
with each instance executing 100 test programs and 140 inputs per program (224k \testcase{}s in total).
For testing InvisiSpec, CleanupSpec, STT\rev{, and SpecLFB} (\cref{sec:eval-def}), we run 100 parallel instances, each executing 200 test programs, and 140 inputs per program (2.8M \testcase{}s in total).

On discovering a violation, we analyze it to identify its root cause following the process in \cref{sec:analyzing-violations}.

\subsection{Testing Baseline Out-of-Order CPU}\label{sec:naive_vs_opt}


\rev{To answer RQ1,} we test the \rev{insecure} baseline O3CPU 
against two contracts: \texttt{CT-SEQ}, which \rev{allows} cache-based leaks on sequential execution, and \texttt{CT-COND}, which additionally \rev{allows} leaks on mispredicted branches.
We run campaigns with \naive{} and \opt{} \rev{using the default trace format (L1D-cache and D-TLB)}.
\cref{table:BaselineTesting} shows the 
execution time, \rev{number of violating test cases} detected, and detection time per violation,
averaged over 16 parallel runs.


\begin{table}[htb]
\centering
\caption{Results of testing the baseline Out-of-Order CPU.
}
\label{table:BaselineTesting}
\small
\begin{tabular}{clccc}
 
 \hline
\textbf{Metric} & \textbf{Contract}&\textbf{Naive}&\textbf{Opt}&\textbf{Ratio}\\ \hline
\multirow{2}{*}{\makecell{\textbf{Time (minutes)}}} &CT-SEQ & 289 & 25 & 11.7x\\
                                &CT-COND& 289 & 33 & 8.7x\\ \hline
\multirow{2}{*}{\makecell{\textbf{Number of}\\\textbf{violations}}} &CT-SEQ & 5.8 & 9.9 & 1.7x\\
                                      &CT-COND& 0 & 0.1  & N/A\\ \hline
\multirow{2}{*}{\makecell{\textbf{Detection time}\\\textbf{(minutes)}}} &CT-SEQ & 49.8 & 2.5 & 19.9x\\ 
                                      &CT-COND& N/A & 330  & N/A\\ \hline

\end{tabular}
\end{table}

\compactpara{Detected Violations.}
\naive{} detected violations against \texttt{CT-SEQ}.
These were due to branch mispredictions, namely instances of Spectre-v1, where a value was leaked speculatively on a mispredicted conditional jump instruction.

\opt{} found violations of both \texttt{CT-SEQ} and \texttt{CT-COND}.
The latter were instances of Spectre-v4, where a store is speculatively bypassed by a younger load to the same address, which reads the value in memory from before the store and leaks it by encoding it in a subsequent load's address.

\compactpara{Execution Time.}
The test campaign with \naive{} took 289 minutes (13 \testcases{} per second).
\opt{} improved on this, taking up to 33 minutes per campaign (114 \testcases{} per second).
This speedup is due to the amortized startup cost of the simulator.
\opt{} sometimes requires additional validation of violations, which explains the difference in time for \texttt{CT-SEQ} and \texttt{CT-COND}.
However, the reduction in startup cost significantly outweighs the extra validation cost, leading to a net speedup of 9--11x.



\compactpara{Detection Time}.
\naive{} detects \texttt{CT-SEQ} violations in less than 1 hour on average, demonstrating that even a basic \Gemfuzzer{} implementation is useful in detecting leaks.
\opt{} detects a \texttt{CT-SEQ} violation in 2.5 minutes (20x faster) and a \texttt{CT-COND} violation in 6.5 hours.
\rev{It} takes longer to discover Spectre-v4 (\texttt{CT-COND} violation) compared to Spectre-v1 (\texttt{CT-SEQ} violation) as the probability that a load and store address match and mispredict is low in random tests.\looseness=-1


\opt{} finds more violations compared to \naive{} with the same number of tests because it initializes the cache with full sets, which results in violations through speculative installs and also via evictions. \opt{} also preserves the branch predictor and memory-dependence predictor state between inputs, which allows a broader set of predictions and increases the chances of a violation. 

As \opt{} performs significantly better,
we only use this version next and we refer to it simply as \Gemfuzzer{}.

\begin{table*}[tb]
\centering
\caption{Results of testing
InvisiSpec, CleanupSpec, STT\rev{, SpecLFB}, and the baseline Out-of-Order CPU with \opt{}.
The campaigns consisted of 100 parallel instances of \Gemfuzzer{}, each executing 200 programs, each with 140 inputs.
}
\small
\begin{tabular}{l|cccc|cc}
\hline
\textbf{Defense} & \textbf{Contract} & \textbf{Detected} & \textbf{Avg. Detection} & \rev{\textbf{Number of}} & \textbf{Testing Throughput} &  \textbf{Campaign}\\
 & & \textbf{Violation?} & \textbf{Time (sec.)} & \rev{\textbf{Unique Violations}} &  \textbf{(\testcases{}/sec.)} & \textbf{Execution Time}\\
\hline
Baseline & CT-SEQ     & YES  & 1.7 &  \rev{2} & 752  & 1 hr 2 min \\ 
\hline
InvisiSpec & CT-SEQ   & YES  & 2.1 & \rev{1} & 630  & 1 hr 14 min \\ %
CleanupSpec & CT-SEQ  & YES  & 1.1 & \rev{3} & 2592 & 18 min \\ 
\rev{SpecLFB}     & \rev{CT-SEQ}  & \rev{YES}  &  \rev{1.6}   & \rev{1} &  \rev{2595} &  \rev{18 min} 
\\\hline
STT & ARCH-SEQ    & YES & 10371 & \rev{1} & 34  & 23 hr 3 min \\\hline

\end{tabular}
\label{table:campaigns}
\end{table*}

\rev{
\subsection{Evaluating different \utrace{} formats}\label{sec:trace_formats}

To answer RQ2 and assess the trade-off between precision, performance, and percentage of violating test cases detected by different \utrace{s}, we ran testing campaigns on the baseline Out-of-Order CPU with four types of \utrace{s} exposing different fine-grained \uarch{} information:

\begin{itemize}[left=0pt]
    \item {\bf Baseline (L1D+TLB):} The \utrace{} consists of the final addresses in the L1D cache and D-TLB.
    \item {\bf BP State:} The \utrace{} consists of the final state of the branch predictor (BP) and the end of the test case.
    \item {\bf Memory access order:} The \utrace{} consists of the ordered list of  all memory accesses (PCs and addresses).
    \item {\bf Branch prediction order:} The \utrace{} consists of the ordered list of branch PCs and their predicted targets.
\end{itemize}

We measure the \textit{fraction of total violations} as the percentage of violations detected by a trace format divided by the total violations detected by any trace format. For each of the formats, we also report the fraction of their violations also detected by baseline trace format. \cref{table:trace_formats} shows the results.


\begin{table}[bht]\rev{
\centering
\caption{\rev{
Results of testing the baseline O3CPU with different \utrace{} formats, across 100 parallel instances of AMuLeT
each executing 200 programs, each with 140 inputs, compared to baseline trace (L1D Cache and TLB).  
}
}
\label{table:trace_formats}

\begin{tabular}{p{1.85cm}|p{1.8cm}|p{1.7cm}|p{1.7cm}}\hline
\small
    {\bf Trace format}&{\bf Throughput (test cases / sec.)}&{\bf Fraction of total violations}&{\bf Violations covered by baseline trace}\\\hline
Baseline (L1D+TLB)&580&79.8\%&100\%\\\hline
BP state&27&6.9\%&70.8\%\\\hline
Memory access order& 67&91.9\%&80.9\%\\\hline
Branch prediction order& 302&2.6\%&77.8\%\\\hline

\end{tabular}}
\end{table}


The BP state, branch prediction order, and memory access order \utrace{s} have lower throughput than the baseline. This is because these 
\uarch{} traces suffer from additional \textit{validations}. Recall from \cref{sec:htrace_design} that violations may be observed due to difference in the initial \uarch{} context rather than inputs, and \tool{} confirms or rejects a violation by re-running the inputs with the same initial \uarch{} state. 
These traces are more impacted by varying initial \uarch{} states, causing more validations and lowering the test throughput.


The baseline \utrace{} consisting of the final L1D cache and D-TLB state detects almost 80\% of the total violating test cases
without significantly slowing down the fuzzer.
Although the memory access order trace detects a higher fraction (92\%) of the violations, it is an order of magnitude slower in throughput due to extra validations, and not all of its violations may be directly exploitable by a realistic attacker.
While we confirm that the BP state or branch prediction order traces detect instances of implicit channels based on branch predictions, a  majority ($>$70\%) of these violations 
are also detected by the baseline \utrace{}, since differences in the speculative control flow may also manifest as differences in data-flow and accessed addresses.

The baseline \utrace{} (L1D cache and D-TLB state) provides the best trade-off between speed and coverage. Since most defenses~\cite{cleanupspec, invisispec, STT, speclfb} protect against leaks through the cache, 
we use the baseline \utrace{} (L1D cache and D-TLB state) in the rest of the evaluation.
This is sufficient to detect leakages in all defenses we test.

} 

\subsection{Testing Defenses with \Gemfuzzer{}}\label{sec:eval-def}

To answer \rev{RQ3} and \rev{RQ4}, we test \rev{four} secure speculation mechanisms, InvisiSpec, CleanupSpec, STT\rev{, and SpecLFB}.
We also test the baseline CPU as a comparison point.
We test InvisiSpec,  CleanupSpec, \rev{and SpecLFB} against \texttt{CT-SEQ} as \rev{these defenses claim to protect against speculative  memory-system side channels}~\cite{invisispec,cleanupspec,speclfb}.
We test STT \rev{with} \texttt{ARCH-SEQ}, which captures STT's non-interference guarantee~\cite{hw_sw_contracts, STT}.
\cref{table:campaigns} shows the results of these campaigns.

\vspace{0.05in}
\noindent \textbf{Types of Violations.} For the baseline non-secure CPU, we find similar violations as previously discussed in \cref{sec:naive_vs_opt}.
We also find contract violations for InvisiSpec, CleanupSpec, STT\rev{, and SpecLFB} and analyze their root causes in \cref{sec:IS-violations}--\cref{sec:stt-violations}.
Surprisingly, a majority of these are new leakages that differ from previously known vulnerabilities ~\cite{unXpec,SpecInterferenceAttack, DOLMA}.

\vspace{0.05in}
\noindent \textbf{Performance.} The violations in InvisiSpec, CleanupSpec\rev{, and SpecLFB} were discovered in less than 3 seconds on average\rev{. This is} comparable to the baseline CPU, whereas on STT, it takes longer (3 hours on average).
CleanupSpec \rev{and SpecFLB} tests are faster than InvisiSpec (\rev{2590+} vs 630 \testcases{}/second) as they start from a clean cache state (which only requires flushes to the sandbox addresses), whereas InvisiSpec requires filling the cache with addresses that conflict with the sandbox, requiring more instructions.


STT tests are much slower (34 \testcases{}/second) as its simulation on Gem5 is 17x slower than InvisiSpec, due to its higher complexity.
The overall testing performance for the Baseline is higher compared to \cref{sec:naive_vs_opt} (752 vs 114 \testcases{}/second for the baseline) because we run 100 parallel instances of \Gemfuzzer{} in this campaign instead of 16.




\subsection{Vulnerabilities Found in InvisiSpec} \label{sec:IS-violations}


InvisiSpec's design~\cite{invisispec} claims that speculative loads are invisible to the caches; when loads become safe, an \textit{expose} operation makes the load visible to the caches (installing and evicting cache lines).
Below, we describe the leaks we discovered in InvisiSpec, differentiating between Known Vulnerabilities (KV) and Unknown Vulnerabilities (UV). 

\smallskip 
\compactpara{UV1. Speculative L1 D-Cache Evictions.} 
All violations in InvisiSpec in \cref{table:campaigns} were due to a previously unknown vulnerability caused by an implementation bug. 
\cref{fig:specL1eviction_InvisiSpec} shows one such test that caused a violation. 
\cref{fig:spec_l1_evictions_asm} (Line 11) shows a mis-speculated load, whose address depends on an input (\texttt{rbx}). 
Based on the input, the speculative load's address differs, as shown in \cref{fig:spec_l1_evictions_logs}. 
We observe that this speculative address is leaked based on an address evicted from the cache. In Input A, the speculatively accessed address is \blue{0x3a00}, which evicts the valid address \blue{0x13a00} (initially in the L1D-cache) and is thus absent in the final state.
Instead, in Input B, the accessed address \green{0x3100} evicts \green{0x13100}, which is absent from the \utrace{}. 

\begin{figure}[h!]
    \centering
    \begin{subfigure}[m]{0.5\columnwidth}
        \centering
        \begin{lstlisting}[style=x86asm, basicstyle=\tiny, label=lst4]
.bb_main.2:
OR byte ptr [R14 + RDX], AL 
LOOPNE .bb_main.3 
JMP .bb_main.exit 

.bb_main.3: # misspeculated
AND BL, 34 
AND RAX, 0b111111111111 
CMOVNBE SI, word ptr [R14 + RAX] 
AND RBX, 0b111111111111
XOR qword ptr [R14 + RBX], RDI
        \end{lstlisting}
         \caption{Test Asm Causing Violation}
         \label{fig:spec_l1_evictions_asm}
    \end{subfigure}
    \hfill
    \begin{subfigure}[m]{0.45\columnwidth}
        \centering
    \begin{tikzpicture}[baseline, node distance=2cm and 1.5cm,>=stealth,auto]
        \node[draw, text width=3.5cm, align=left] (node1) {
         \scriptsize
            \textbf{Input A} \\
            ... 0x130c0 \green{0x13100} ... 0x139c0 ...
            \textbf{Input B} \\
            ... 0x130c0 ... 0x139c0 \blue{0x13a00} ...\\
        };

        \node[draw, text width=3.5cm, align=left,below of=node1] (node2) {
         \scriptsize
            \textbf{Input A} \\
            Load Line-11, Addr: \blue{0x3a10}\\
            \textbf{Input B} \\
            Load Line-11, Addr: \green{0x3110}\\
        };

        \node[anchor=north] at ($(node1.north) + (0,0.5)$) {\scriptsize \textbf{L1D-cache Tags in \utrace{}}};
         \node[anchor=north] at ($(node2.north) + (0,0.5)$) {\scriptsize \textbf{Speculative Load in Program}};
        \end{tikzpicture}
    \caption{Two inputs to Asm that evict different addresses}
    \label{fig:spec_l1_evictions_logs}
    \end{subfigure}
    \caption{Example of Test Asm Causing Violation in InvisiSpec due to Speculative L1D-cache Evictions.}
    \label{fig:specL1eviction_InvisiSpec}
\end{figure}
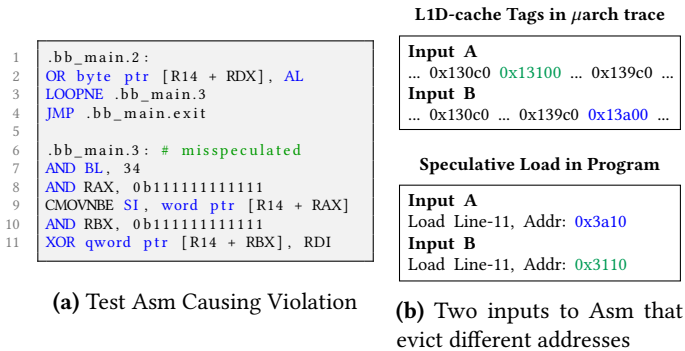

\begin{figure}[h!]
\vspace{-0.1in}

\begin{lstlisting}[style=CStyle, caption=InvisiSpec's Speculative L1-Cache Eviction Bug, basicstyle=\tiny, label=lst:spec-eviction-bug]
    // Cache Miss
    if (L1Dcache.cacheAvail(in_msg.LineAddress)) {
        // Space Available, Send Request to L2
        trigger(mandatory_request_type_to_event
        (in_msg.Type), in_msg.LineAddress, ...);
        
    } else { // No Space Available
        // L1 Eviction (Even For Speculative Requests)
        trigger(Event:L1_Replacement, 
        L1Dcache.cacheProbe(in_msg.LineAddress), ...);
    }
\end{lstlisting}

\begin{lstlisting}[style=CStyle, caption=Patch for InvisiSpec Speculative Eviction Issue, basicstyle=\tiny, label=lst:spec-eviction-fix]
    if (L1Dcache.cacheAvail(in_msg.LineAddress) 
        @|| (in_msg.Type == RubyRequestType:SPEC_LD)@) {
        // Space Available or Spec Load, Send Request to L2
        trigger(mandatory_request_type_to_event
        (in_msg.Type), in_msg.LineAddress, ...);
        
    } else { // L1 Eviction Only for Non-Speculative Requests
        trigger(Event:L1_Replacement, ...
\end{lstlisting}
\caption{Bug in InvisiSpec's Gem5 implementation discovered by \Gemfuzzer{} that leaks addresses of speculative loads via L1D-cache evictions and breaks its security guarantees.}
\end{figure}

\smallskip\noindent \textit{Root Cause:} While speculative loads in InvisiSpec are supposed to be invisible to the cache hierarchy, we observe that its Gem5 implementation does not match this.
As shown in the code snippet from InvisiSpec in \cref{lst:spec-eviction-bug} (line 9), on a speculative load causing a L1D-cache miss, 
if a cache set is fully occupied, we observe that a speculative load initiates an L1D-cache replacement regardless of whether it is safe or not to do so.
So,  a mis-speculated load might evict a conflicting address from the L1D-cache, thus leaking its address and breaking InvisiSpec's security guarantees.

\smallskip\noindent \textit{Fix:} This issue is an implementation bug and can be easily patched. As shown in \cref{lst:spec-eviction-fix}, we modify the implementation so that L1-replacements are only executed on non-speculative loads, i.e., when the load is safe and ready to be exposed to the cache hierarchy. 
This fixes the leakage, as campaigns after the patch in ~\cref{sec:results_amplify_vio} found no violations.


\smallskip
\compactpara{KV1. Speculative Instruction Fetches.} 
In previous campaigns, when we included the L1I-cache in the \utrace{}, we detected violations where L1I-cache state differed between two inputs. This is because, based on an input, the execution time can vary due to differences in speculative hits and misses, causing differences in instruction fetch behavior that speculatively brings additional lines into the L1I-cache. 
InvisiSpec~\cite{invisispec} acknowledges that it does not protect the L1I-cache, making this a known vulnerability.
This shows \Gemfuzzer{}'s ability to detect unprotected threat vectors.

\subsubsection{Amplifying the Vulnerability in InvisiSpec} \label{sec:results_amplify_vio}

\begin{table}[htb]
\centering
\caption{Results of testing InvisiSpec (Patched) with smaller \uarch{} structures (fewer L1D-cache ways and MSHRs)}
\small
\begin{tabular}{ccc}
\hline
\textbf{InvisiSpec Configuration} & \textbf{Time} & \rev{\textbf{Violation}}\\\hline
Patched, 8-way L1D, 256 MSHRs &  52 min & \rev{\xmark} \\
Patched, 2-way L1D, 256 MSHRs  & 20 min & \rev{\xmark} \\
Patched, 2-way L1D, 2 MSHRs  & 26 min & \rev{\cmark} \\\hline
\end{tabular}
\label{table:amplify_invisispec}
\end{table}

We add the bug-fix in \cref{lst:spec-eviction-fix} and continue our testing campaigns of InvisiSpec (Patched), running 100 parallel runs of 200 tests.
As shown in \cref{table:amplify_invisispec}, we do not observe further violations after patching with the default, 8-way L1D-cache.
To answer \rev{RQ3} and see if we can amplify the vulnerability, we test with smaller \uarch{} structures that experience \rev{higher} contention.
We run campaigns with 2-way L1D-cache (down from 8 ways) and with D-cache Miss-Status-Handling Registers (MSHRs) reduced to 2 (\rev{down} from the default 256).

\cref{table:amplify_invisispec} shows that reducing the L1D-cache from 8 to 2 ways speeds up fuzzing campaigns by 2.6$\times$. This is because initializing a smaller L1D-cache requires fewer instructions per test case. But this does not result in new violations.
However, reducing MSHRs from 256 to 2 reveals new violations. 
These leaks stem from a previously unknown variant of the \textit{speculative-interference attack}~\cite{SpecInterferenceAttack}.
Unlike prior attacks, which requires a multi-threaded attacker and SMT, our new variant (\textbf{UV2}) allows attacker observations from the same core, breaking InvisiSpec's secuity in a single-threaded setting.

\begin{figure}[h!]
    \centering
    \begin{subfigure}[m]{0.5\columnwidth}
        \centering
        \begin{lstlisting}[style=x86asm, basicstyle=\tiny, label=lst5]
.bb_main.1:
AND DL, 7
AND RSI, 0b111111111111
CMOVL EAX, dword ptr [R14 + RSI] 
...
OR qword ptr [R14 + RCX], RAX 
JNO .bb_main.2 
JMP .bb_main.exit 

.bb_main.2: # misspeculated
AND RCX, 0b111111111111
LOCK AND dword ptr [R14 + RCX], EDI 
AND RAX, 0b111111111111
OR EDI, dword ptr [R14 + RAX]
        \end{lstlisting}
         \caption{Test Asm Causing Violation}
         \label{fig:spec_interference_asm}
    \end{subfigure}
    \hfill
    \begin{subfigure}[m]{0.45\columnwidth}
        \centering
    \begin{tikzpicture}[baseline, node distance=2cm and 1.5cm,>=stealth,auto]
        \node[draw, text width=3.5cm, align=left] (node1) {
         \scriptsize
            \textbf{Input A} \\
            ... 0x38c0 0x3f40 ... \\
            \textbf{Input B} \\
            ... 0x38c0 \red{0x3e80} 0x3f40 ...
            
        };

        \node[draw, text width=3.5cm, align=left,below of=node1] (node2) {
         \scriptsize
            \textbf{Input A} \\
            PC: 0x402152, Addr: \blue{0x3500}\\
            \textbf{Input B} \\
            PC: 0x402152, Addr: \green{0x3140}\\
        };

        \node[anchor=north] at ($(node1.north) + (0,0.5)$) {\scriptsize \textbf{L1D-cache Tags in \utrace{}}};
         \node[anchor=north] at ($(node2.north) + (0,0.5)$) {\scriptsize \textbf{Speculative Load in Program}};
        \end{tikzpicture}
    \caption{Two inputs to Asm that speculatively access different addresses}
    \label{fig:spec_interference_logs}
    \end{subfigure}
    \caption{Test Asm Causing Violation in InvisiSpec due to Speculative Interference because of MSHR contention.}
    \label{fig:spec_interference_InvisiSpec}
\end{figure}
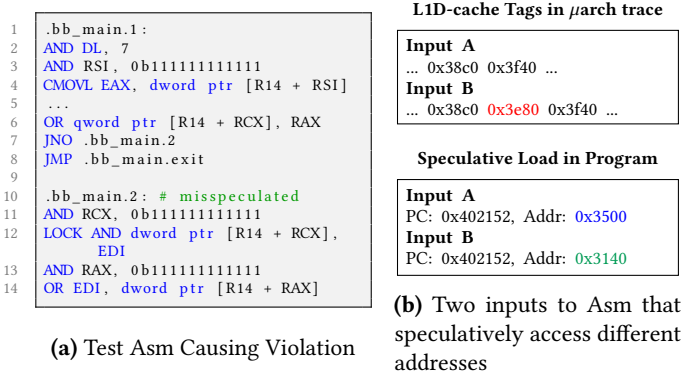

\compactpara{UV2. Same-Core Speculative Interference Attack.} In the example generated by \Gemfuzzer{}, shown in \cref{fig:spec_interference_asm}, we have a speculative load, \textbf{SL} (Line 14 in \cref{fig:spec_interference_asm}), which shares the MSHRs with another non-speculative load, \textbf{NSL} (Line 6). Based on whether \textbf{SL} has a cache hit/miss, it causes contention on the MSHRs, subsequently delaying the execution of \textbf{NSL}. 
As shown in \cref{table:ipc-mshr-contention}, Input A has an \textbf{SL} to address \blue{0x3500},
which misses the L2 cache, so an MSHR is occupied by this speculative request; this
stalls a later Expose operation of the \textbf{NSL} to \red{0x3e80}. 
Since InvisiSpec must perform an Expose operation to move lines in its speculative buffer to the cache, and \red{0x3e80} is unable to issue its Expose request before the test case ends, it is never brought into cache.
Conversely, Input B has a \textbf{SL} to \green{0x3140}, which hits in the L2 cache;
so the occupied MSHR is freed up quickly, allowing the Expose of the \textbf{NSL} to \red{0x3e80} to issue and be installed in the L1 cache before the test case ends.
By accessing \red{0x3e80} after the test ends, an adversary can observe whether the access was fast or slow, dependent on input A or B.

\vspace{0.1in}
\noindent
\textit{Root cause:} The delayed \textbf{NSL} Expose is in fact observable by an adversary that subsequently accesses \textit{any} address.
As the cache-controller queues are in-order, the stall due to the Expose at the head of the queue causes \textit{all} subsequent requests (e.g., a cache hit to \texttt{0xbeef}) to be stalled; whereas in the case where the \textbf{NSL} is not delayed, the request (e.g., a cache hit to \texttt{0xbeef}) is serviced faster.  
We also discovered variants of this leakage where MSHR contention is due to a miss in InvisiSpec's speculative buffer (instead of L2 cache). 

\vspace{0.1in}
\noindent
\textit{Fix:} This vulnerability is a fundamental design issue. 
A redesign that addresses the original speculative interference attacks, such as GhostMinion~\cite{ghostminion}, which ensures that younger loads cannot influence the execution time of older loads, will also address the variants we discovered. 

\begin{table} [h!]
\caption{Explanation of MSHR interference violation in InvisiSpec.
The column MSHRs show the addresses in the L1-MSHRs. \blue{Blue} and \green{Green} are speculative loads (SL), while \red{Red} and Black are non-speculative loads (NSL). Input A induces speculative MSHR interference, but Input B does not.}

\renewcommand{\arraystretch}{1.15} 
\fontsize{8pt}{8pt}\selectfont
\begin{tabular}{|c|c|c|c|}
    \hline
    \multicolumn{2}{|c}{\textbf{Input A}} & \multicolumn{2}{|c|}{\textbf{Input B}} \\\hline
    \textbf{Operation} & \textbf{MSHRs}
    & \textbf{Operation} & \textbf{MSHRs} \\\hline
    \blue{SpecLoad 0x3500} & \blue{3500} & \green{SpecLoad 0x3140} & \green{3140} \\\hline
    --- & \blue{3500} &
    \green{L2 hit 0x3140} & \\\hline
    Replace 0x13e80 & \blue{3500}, 13e80 & Replace 0x13e80 & 13e80 \\\hline
    \red{Expose 0x3e80-- \textbf{stall!}} & \blue{3500}, 13e80 & \red{Expose 0x3e80} & 13e80, \red{3e80} \\\hline
    m5exit & \blue{3500}, 13e80 & m5exit &
    13e80, \red{3e80} \\\hline
    Load 0xbeef-- \textbf{slow!} & \blue{3500}, 13e80 & Load 0xbeef-- \textbf{fast!} & 13e80, \red{3e80} \\\hline
\end{tabular}
\label{table:ipc-mshr-contention}
\end{table}

\subsection{Vulnerabilities Found in CleanupSpec} \label{sec:CS-violations}
CleanupSpec~\cite{cleanupspec} allows speculative loads to modify cache state \rev{and cleans up these state changes on a mis-speculation.}
Below, we describe the leaks we discovered in CleanupSpec and classify them as per their root cause in \cref{table:cleanupspec_breakdown}. We identify 2 new bugs causing insufficient cleaning 
and 1 new vulnerability causing excessive cleaning.

\ignore{
In our testing, we discovered violations due to leaks in CleanupSpec
We analyze the violations we found in CleanupSpec and classify them based on their root cause, in \cref{table:cleanupspec_breakdown}.
We identify three root causes: 2 new bugs resulting in insufficient cleaning of speculative cache state, and 1 new vulnerability, resulting in excessive cleaning of the cache state. 
}



\begin{table}[bht]
\centering
\small
\caption{\rev{Types of} CleanupSpec violations found from 100 parallel runs of 200 test programs, 140 inputs, with the unmodified CleanupSpec (Original) and after a fix for speculative-stores not being cleaned up (Patched).}
\begin{tabular}{lcc}
\hline 
\textbf{\rev{Violation Type}}&\textbf{Original} &\textbf{Patched}\\ \hline 
Speculative Store Not Cleaned   & \rev{\cmark} & \rev{\xmark}\\
Split Requests Not Cleaned      & \rev{\cmark}  & \rev{\cmark}\\
Too Much Cleaning               & \rev{\cmark}  & \rev{\cmark}\\
\hline

\end{tabular}
\label{table:cleanupspec_breakdown}
\end{table}

\compactpara{UV3. Speculative Store Not Cleaned Bug.}
\rev{The first violation in CleanupSpec was due to incorrect cleanup for speculative stores.}
As shown in \cref{lst:cleanupspec-bug-specstore},
cleaning speculative cache state requires tracking cache hit/miss metadata.
While this tracking exists in \texttt{readCallback()} \rev{for} speculative loads, it is missing \rev{in} \texttt{writeCallback()} \rev{for speculative stores}. 
Consequently, addresses of speculative stores
are not cleaned on a mis-speculation, causing leakage.
We discovered this when Spectre-v1-like tests with speculative stores caused L1D-cache differences, and the debug log showed incorrect metadata, as shown in  \cref{fig:cleanupspec_bugs}.
We patch\rev{ed} this by \rev{fixing} \texttt{writeCallback()} to update cleanup metadata on speculative stores. After the fix (\textbf{Patched} in \cref{table:cleanupspec_breakdown}), these violations no longer occur, resolving the bug. 

\begin{figure}[h!]
\vspace{-0.1in}
    \centering
    \begin{subfigure}[m]{0.45\columnwidth}
        \centering
    \begin{lstlisting}[style=CStyle_NoLinenum, caption=CleanupSpec bug - speculative store not cleaned, basicstyle=\tiny, label=lst:cleanupspec-bug-specstore]
// Logic in ReadCallback()
if (L1Hit)
    pkt->setL1Hit();
...    
// Metadata for Cleaning
if (!pkt->isL1Hit())
    DO_CLEANUP(...);
        \end{lstlisting}
        
\begin{tikzpicture}
    \centering
     \node[fill=gray!10, draw=black, align=left,text width=3.65cm,xshift=-10pt](node1){
      \scriptsize
         WriteCallback() for Spec Store \\
         \red{L1Hit: 0, L2Hit: 0, L2Miss: 0}\\
     };
    \node[anchor=north] at ($(node1.north) + (0,0.5)$) {\scriptsize \textbf{Debug Log on Speculative Store}};     
      \end{tikzpicture}
\end{subfigure}
    \hfill
    \begin{subfigure}[m]{0.47\columnwidth}
        \centering
        \begin{lstlisting}[style=CStyle_NoLinenum, caption=CleanupSpec Bug: no cleanup for split request, basicstyle=\tiny, label=lst:cleanupspec-bug-split]

if (loadQueue[load_idx]->isSplitReq()){

    ++lsqSquashedLoadsSplitReq;
    
    // TODO: Cleanup for SplitReq
}
       \end{lstlisting}
  \end{subfigure}
   \caption{Implementation bugs in CleanupSpec's gem5 implementation discovered by \Gemfuzzer{}'s testing campaigns.}
   \label{fig:cleanupspec_bugs}
\end{figure}

\vspace{0.05in}
\compactpara{UV4. Split Requests Not Cleaned Bug.} 
\rev{We discovered a subsequent violation} due to a \rev{CleanupSpec} bug \rev{related to cleanup of requests crossing cache line boundaries}. 
When a load or store accesses data across cache lines (e.g., a 4-byte store to an address two bytes from the cacheline boundary), they spawn split-requests in gem5, i.e. multiple memory-system requests.
We see that CleanupSpec does not clean speculative split requests, as shown in \cref{lst:cleanupspec-bug-split}
from CleanupSpec code. 
We detected this with test cases like Spectre-v1 where speculative loads crossed cacheline boundaries.

\vspace{0.05in}
\compactpara{UV5. Too Much Cleaning Vulnerability.} 
Finally, we discover a new vulnerability in CleanupSpec due to \rev{an incorrect cleanup} of non-speculative \rev{loads}
when non-speculative loads \rev{get} reordered with transient loads to the same cache line.


\cref{table:too_much_cleaning} shows the operations in a test causing such a violation.
Input A has a younger, speculative load \textbf{SL} (\texttt{PC = 0x40114d}) to the same address as an older non-speculative load \textbf{NSL} (\texttt{PC = 0x40113c}). 
For input B, the \textbf{SL} address differs from the \textbf{NSL} address.
In both cases, the \textbf{SL} has a cache miss, installs the address, and is cleaned and evicted on a mis-speculation. 
However, for Input A, cleaning also removes any trace of the \textbf{NSL} to that address. 
In contrast, for Input B, as the \textbf{NSL} address differs from the \textbf{SL}, it remains in the cache after mis-speculation. 
Thus, these two inputs have different \utrace{}s resulting in leakage. 

\smallskip\noindent\textit{Root cause and Fix:} This is the first discovered speculative interference vulnerability~\cite{SpecInterferenceAttack} on CleanupSpec, where interaction between transient and non-speculative loads corrupts cleaning metadata. 
A potential mitigation \rev{can} identify the reordering of non-speculative \rev{and} transient loads to the same addresses at commit time, and set a \texttt{noClean} flag for \rev{the} younger speculative load(s).
We leave \rev{this} for future work.




\begin{table}[h!]
\centering
\caption{Test cases showing ``Too Much Cleaning'' vulnerability in CleanupSpec: sequence of operations}
\footnotesize
\setlength{\tabcolsep}{4pt}

\begin{tabular}{|m{0.6cm}|m{1cm}|m{0.75cm}|m{0.7cm}|m{0.6cm}|m{1cm}|m{0.75cm}|m{0.7cm}|}
\hline
\multicolumn{4}{|c|}{\textbf{Input A}} & \multicolumn{4}{c|}{\textbf{Input B}} \\ \hline
\textbf{Cycle} & \textbf{PC}  & \textbf{Type} & \textbf{Addr.} & \textbf{Cycle} & \textbf{PC} & \textbf{Type} & \textbf{Addr.} \\ \hline 
1060 & 0x4011bb  & Load & 0x1100 & 1060 & 0x4011bb & Load & 0x1100 \\ \hline
1150 & 0x40114d  & SpecLd & \textcolor{red}{0x10c0} & 1150 & 0x40114d & SpecLd & \textcolor{red}{0x1080} \\ \hline
1423 & 0x40113c  & Load & \blue{0x10c0} & 1423 & 0x40113c & Load & \blue{0x10c0} \\ \hline
\textcolor{black}{1429} & 0x40114d  & Undo & \textcolor{red}{0x10c0} & \textcolor{black}{1580} & 0x40114d & Undo & \textcolor{red}{0x1080} \\ \hline
\multicolumn{2}{|c|}{\textbf{\utrace{}}} & \multicolumn{2}{c|}{ 0x1100} & 
\multicolumn{2}{c|}{\textbf{\utrace{}}} & \multicolumn{2}{c|}{0x1100 $~~~$ \blue{0x10c0}} \\\hline

\end{tabular}
\label{table:too_much_cleaning}
\end{table}

\compactpara{KV2. UnXpec~\cite{unXpec} Vulnerability.} 
In campaigns where we included the L1I-cache state in the \utrace{}, we detected violations similar to the unXpec vulnerability~\cite{unXpec}.
In these \rev{violations, the} inputs had different L1I-cache states at the end.
The root cause was 
a difference in cleanup operations causing varying execution times 
\rev{and the instruction fetch speculatively installing} extra lines in the L1I-cache. 

\cref{table:unxspec} shows the operations in one such test case. 
%
Input A has a speculative load to \texttt{0x1180} that is an L1 hit, requiring no cleanup. Whereas, for Input B, the speculative load to \texttt{0x1840} is an L1 miss, that requires a cleanup on mis-speculation.
As cleanup is on the critical path of execution,
this increases the execution time for Input B compared to Input A.
This is like the UnXpec~\cite{unXpec} vulnerability that uses the difference in cleaning operation times to leak information. 

As the execution is slower for Input B, the fetch unit speculatively fetches addresses beyond the end of the test, installing these in the L1I-cache. This causes a violation compared with Input A, which does not exhibit such behavior.


\begin{table}[h!]
\centering
\caption{CleanupSpec vulnerability to UnXpec~\cite{unXpec}}
\footnotesize
\setlength{\tabcolsep}{4pt}
\begin{tabular}{|m{0.6cm}|m{1cm}|m{0.75cm}|m{0.7cm}|m{0.6cm}|m{1cm}|m{0.75cm}|m{0.7cm}|}
\hline
\multicolumn{4}{|c|}{\textbf{Input A}} & \multicolumn{4}{|c|}{\textbf{Input B}} \\ \hline
\textbf{Cycle} & \textbf{PC}  & \textbf{Type} & \textbf{Addr.} & \textbf{Cycle} & \textbf{PC} & \textbf{Type} & \textbf{Addr.} \\ \hline 
1054 & 0x40119b & Load & 0x1180 & 1054 & 0x40119b & Load & 0x1180 \\ \hline
1056 & 0x4011b7 & SpecLd & \textcolor{red}{0x1180} & 1056 & 0x4011b7 & SpecLd & \textcolor{red}{0x1840} \\ \hline
1207 & 0x4011a6 & Load & 0x1940 & 1207 & 0x4011a6 & Load & 0x1940 \\ \hline
&&&& \textcolor{red}{1213} & \textcolor{red}{0x4011b7} & \textcolor{red}{Undo} & \textcolor{red}{0x1840} \\ \hline
\textcolor{blue}{1219} & 0x401190  & Store & 0x1940 & \textcolor{blue}{1240} & 0x401190 & Store & 0x1940 \\ \hline
\end{tabular}
\label{table:unxspec}
\end{table}

\begin{figure}[h!]
    \centering
    \begin{subfigure}[m]{0.5\columnwidth}
        \centering
        \caption{Root Cause of SpecLFB Leak}
        \begin{lstlisting}[style=CStyle_NoLinenum, 
%                          caption=SpecLFB bug: \\ isReallyUnsafe(), 
                          basicstyle=\tiny, 
                          label=lst:speclfb-bug-isreallyUnsafe]
if (isPrevNoUnsafe()):
    // no prior unsafe loads
    clearReallyUnsafe();
else if (!isReallyUnsafe()):
    // prior unsafe loads exist
    setReallyUnsafe();

// only unsafe loads stalled
if (!isUnsafe()):
    clearStall();
        \end{lstlisting}
        \begin{tikzpicture}
            \node[fill=gray!10, draw=black, align=left, text width=7.85cm, xshift=-10pt](node1){
                \scriptsize
                \textcolor{keywordcolor}{bool} \textcolor{functioncolor}{isUnsafe}(): \\
                    \ \ \textcolor{keywordcolor}{return} !\textcolor{black}{isSquashed}() \&\& ( (\textcolor{functioncolor}{isCUSL}() || \textcolor{functioncolor}{isMUSL}()) \&\& \textcolor{red}{isReallyUnsafe}()); \\
            };
            \node[anchor=north] at ($(node1.north) + (0,0.5)$) 
            {\footnotesize \textbf{(c) isUnsafe():} uses isReallyUnsafe (cleared for first speculative load)};
        \end{tikzpicture}
    \end{subfigure}
    \hfill
    \begin{subfigure}[m]{0.38\columnwidth}
        \vspace{-3em}  
        \centering

        \caption{Violation Test Asm}
        \begin{lstlisting}[style=x86asm, basicstyle=\tiny, label=lst:speclfb_spectre_v1_asm]
# RBX is secret
CMP RAX, 0 # non-zero RAX
JNE .l1  
# RAX == 0, misprediction
  MOV RAX, ptr[R14 + RBX]
  JMP .l2
.l1:  
# RAX == 0
  MOV RAX, ptr[R14 + 64]
.l2: ..
  \end{lstlisting}

    \end{subfigure}
    \caption{\rev{Vulnerability in SpecLFB discovered by \tool{}. The root cause is an undocumented optimization in SpecLFB code that clears protections for the first speculative load.}}
    \label{fig:speclfb_bugs}
\end{figure}
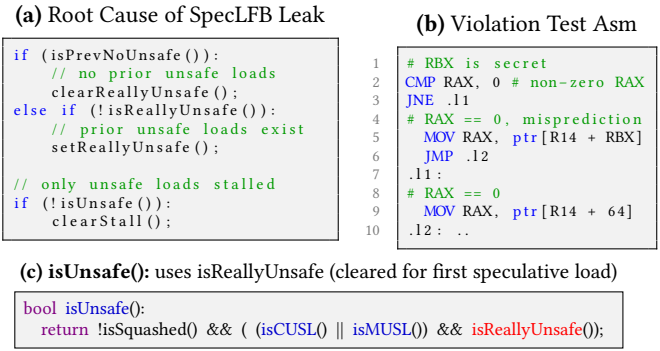

\rev{
\subsection{Vulnerabilities found in SpecLFB}\label{sec:SpecLFB-violations}
SpecLFB~\cite{speclfb} adds security checks to the cache Line-Fill Buffer (LFB)  to prevent leaks via speculative cache misses. It blocks speculative cache misses
from being installed into the cache until they are \textit{safe} (like Delay-On-Miss~\cite{delay-on-miss}). Based on this, we tested SpecLFB's \sheprev{gem5 implementation} against the \texttt{CT-SEQ} contract, \sheprev{the most intuitive contract based on the security guarantees described in the SpecLFB paper~\cite{speclfb},} and surprisingly found \sheprev{it to be insecure with respect to this contract}.


\vspace{0.05in}
\noindent\textbf{UV6. First Speculative Load Not Protected in SpecLFB.} The violating tests discovered by \tool{} on SpecLFB are all similar to Spectre-v1, where the secret is in a register, as shown in \cref{fig:speclfb_bugs}~(b).
We discovered that the root cause of these violations is an undocumented optimization in \sheprev{SpecLFB's implementation} that marks speculative loads incorrectly as \textit{safe} if they are the first speculative load in the load-store queue. 
SpecLFB identifies \textit{unsafe} speculative loads to be delayed by checking a flag, \texttt{isUnsafe}, for each load. The \texttt{isUnsafe} flag depends on whether the load is speculative either due to branch speculation (\texttt{isCUSL}) or memory dependence speculation (\texttt{isMUSL}), \textit{and} it also depends on an \texttt{isReallyUnsafe} flag  that is cleared if the load is the first speculative load, as shown in \cref{fig:speclfb_bugs}~(a,c). Thus, Spectre variants with a single speculative load still leak information as they are not blocked from installing into the cache. We validate that these violations disappear when we expose this leakage of register values in the contract (\texttt{ARCH-SEQ}).

}



\subsection{Vulnerabilities Found in STT}\label{sec:stt-violations}
STT~\cite{STT} uses taints to keep track of registers holding data accessed  speculatively from memory or its derivatives, and it blocks the execution of instructions that leak tainted data via side channels.
Below, we describe the violation (\textbf{KV3}) we discovered in STT, which was due to tainted speculative stores incorrectly \rev{executing and} accessing the TLB.


\smallskip
\noindent\textbf{KV3. Speculative Store Leaking via TLB.} 
The violations in STT \rev{manifested as} a difference in the TLB state in the \utrace{}.
The root cause was a speculative store with tainted address \rev{incorrectly} installing an entry in the D-TLB and leaking its address.
This leak was previously \rev{known}~\cite{DOLMA}.

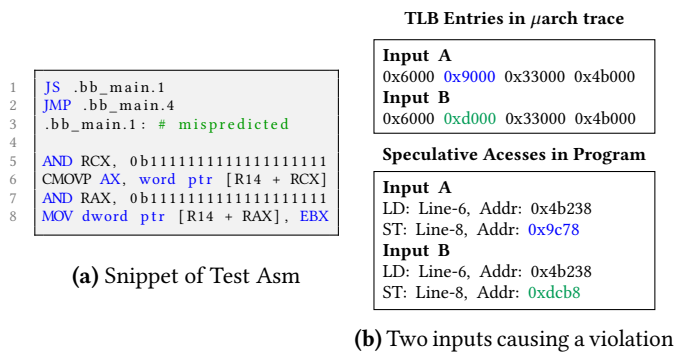
\begin{figure}[h!]
    \centering
    \begin{subfigure}[m]{0.45\columnwidth}
        \centering
        \begin{lstlisting}[style=x86asm, basicstyle=\tiny, label=lst6]
JS .bb_main.1 
JMP .bb_main.4 
.bb_main.1: # mispredicted

AND RCX, 0b1111111111111111111
CMOVP AX, word ptr [R14 + RCX] 
AND RAX, 0b1111111111111111111
MOV dword ptr [R14 + RAX], EBX 
        \end{lstlisting}
         \caption{Snippet of Test Asm}
    \end{subfigure}
    \hfill
    \begin{subfigure}[m]{0.5\columnwidth}
        \centering
    \begin{tikzpicture}[baseline, node distance=2cm and 1.5cm,>=stealth,auto]
        \node[draw, text width=3.5cm, align=left] (node1) {
         \scriptsize
            \textbf{Input A} \\
            0x6000 \blue{0x9000} 0x33000 0x4b000 \\ 
            \textbf{Input B} \\
            0x6000 \green{0xd000} 0x33000 0x4b000 \\ 
        };

        \node[draw, text width=3.5cm, align=left,below of=node1,node distance=2.05cm] (node2) {
         \scriptsize
            \textbf{Input A} \\
            LD: Line-6, Addr: 0x4b238\\
            ST: Line-8, Addr: \blue{0x9c78}\\
            \textbf{Input B} \\
            LD: Line-6, Addr: 0x4b238\\
            ST: Line-8, Addr: \green{0xdcb8}\\
        };

        \node[anchor=north] at ($(node1.north) + (0,0.5)$) {\scriptsize \textbf{TLB Entries in \utrace{}}};
         \node[anchor=north] at ($(node2.north) + (0,0.45)$) {\scriptsize \textbf{Speculative Acesses in Program}};
        \end{tikzpicture}
    \caption{Two inputs causing a violation} 
    \end{subfigure}
    \vspace{-0.1in}
    \caption{Example of Test Asm Causing Violation in STT due to speculative store installing TLB entry.}
    \vspace{-0.15in}
    \label{fig:tlb_spec_store}
\end{figure}

\cref{fig:tlb_spec_store} shows an example violation.
In \cref{fig:tlb_spec_store}(a), 
a  mispredicted conditional jump (\texttt{JS}) in Line 1 causes \texttt{CMOVP} to speculatively load data in Line 6, which is encoded in the store address in Line 8.
\cref{fig:tlb_spec_store}(b) shows two inputs (A and B) leading to a violation. In both, speculatively loaded data is encoded in the store address (\blue{0x9c78} in A and \green{0xdcb8} in B).
While these do not appear in the cache, their TLB entries appear in the \utrace{}, leaking the speculatively accessed data.
The root cause is an implementation bug that allows tainted speculative stores to be executed; blocking their TLB access, as proposed by DOLMA~\cite{DOLMA}, would fix this leak.

\rev{
\section{Discussion}

\subsection{Portability to Other Defenses, Simulators, ISAs}
Due to \Gemfuzzer{}'s modular structure consisting of the (1) test case \textit{generator}, (2) the \textit{leakage model} that provides contract trace, and (3) the \textit{executor} that provides the \utrace{}s, it is quite easy to port \Gemfuzzer{} to different defenses simply by changing the executor.
\cref{table:loc} shows the lines of code (LoC) added or modified for each defense we tested.
A majority of the LoC added/modified in the Gem5 simulator are for the test orchestration (452-532) and inter-process communication with \tool{} (270-353), which are isolated and largely independent of the defense or simulator. The LoC for trace extraction (226-319) added to different \uarch{} components in the simulator is minimal.
The majority of these LoC can be copied from one defense to another with minor changes.

\begin{table}[htb]
\rev{
\centering
\caption{\rev{Lines of Code (LOC) added/changed to different defenses to enable testing with \Gemfuzzer{}.}}
\label{table:loc}
\vspace{-0.1in}
\footnotesize
\begin{tabular}{c|ccc|c}
\hline
\textbf{Defense} &  \textbf{Test} & \textbf{Socket-Based} & \textbf{Trace} &  \textbf{Total}\\ 
& \textbf{Harness} & \textbf{Communication} & \textbf{Extraction} & \textbf{LoC}\\
\hline
InvisiSpec & 532 & 353 & 445 & 1330  \\
CleanupSpec & 452 & 270 & 226 & 948 \\
STT & 486 & 276 & 296 & 1058 \\
SpecLFB & 465 & 338 & 319 & 1122 \\\hline

\end{tabular}
}
\end{table}

\Gemfuzzer{} is easily portable to any CPU simulator, as long as it models CPU speculation (so it exhibits speculation-based vulnerabilities), and allows examining, extracting and resetting the \uarch{} state (cache, branch predictor states, etc.).
This covers most common \uarch{} simulators (e.g., Gem5~\cite{gem5}, Marss-x86~\cite{Marss}, Champsim~\cite{Champsim}).

Porting \Gemfuzzer{} to other ISAs (ARM, RISC-V) is also feasible, but it requires a test case generator that support the new ISA semantics and requires support for accurate emulation of the new ISA in Unicorn~\cite{Unicorn} used by the leakage model. For example, while porting \Gemfuzzer{} to Ghostminion~\cite{ghostminion} (built on the ARM ISA), we discovered bugs in Unicorn's ARM hooks that hindered our testing.
Future work can address this and extend \Gemfuzzer{} to other ISAs.


\subsection{Detecting Additional Leaks with \Gemfuzzer{}}
Our campaigns using \Gemfuzzer{} primarily test for leaks through caches and TLB, because most countermeasures~\cite{invisispec,cleanupspec,speclfb} protect against cache-based leakage. 
However, by including additional \uarch{} state in the \utrace{}s (e.g., branch-predictor state, execution order of loads or branches), \Gemfuzzer{} can detect a wider range of speculative leaks, as shown in \cref{table:trace_formats}. 
Additionally, the state of other \uarch{} predictors, like cache way-predictors, value predictors, or prefetchers, can be added to the \utrace{} to  discover new vulnerabilities~\cite{GoFetch, PandorasBox}.
Thus, \Gemfuzzer{} can be used to 
detect leaks from new micro-architectural features to be added in to a processor well before it is commercialized.

\subsection{Limitations of \Gemfuzzer{}}
\textbf{Limits of Randomized Testing.}
AMuLeT, like any randomized testing tool, can only demonstrate insecurity, but cannot  prove a defense's security.\footnote{~\rev{``Testing shows the presence, not the absence of bugs.'' - Djikstra ~\cite{djikstraTesting}}} 
This makes it complementary to, not a replacement for, formal verification techniques, which can prove security but often lack scalability.
For instance, Pensieve~\cite{pensieve} requires manual modeling of defenses in domain-specific languages and several hours to verify security for programs up to nine cycles.
In contrast, \Gemfuzzer{} finds vulnerabilities in seconds and is easily portable to new defenses, making it a cost-effective first step to identify insecurity (e.g., how we show SpecLFB~\cite{speclfb}'s \sheprev{implementation} is insecure) before attempting complex formal verification. 




\smallskip{}
\noindent \textbf{Limited Search Space.}
\Gemfuzzer{}, like prior black-box fuzzing works~\cite{revizor,HideSeekSpectre,hofmann2023SpecAtFault}, is limited in the search space it can test within the vast \uarch{} state space and design space possible for a processor.
\Gemfuzzer{} focuses its search by reducing the sizes of \uarch{} structures (e.g., reducing MSHRs and cache ways) to prioritize  high-contention states where leaks are more likely.
This prioritization increases the likelihood of finding exploitable speculative leaks without altering the simulator or defenses under test. 
However, the extensive state space remains a limitation for comprehensive leak discovery, and future work may explore adaptive strategies for coverage-guided fuzzing or focused test generation  to improve AMuLeT’s detection capabilities.
}
\section{Related Work}

\smallskip
\noindent
\textbf{Post-silicon fuzzing for microarchitectural leaks.}
\textit{Covert Shotgun}~\cite{covert_shotgun} and \textit{ABSynthe}~\cite{gras2020absynthe} automatically test CPUs for covert channels emanating from \uarch{} contention. 
\textit{Osiris}~\cite{osiris} uses fuzzing to discover new side channels. 
\textit{Transynther}~\cite{medusa} tests CPUs for new variants of MDS attacks by mutating  attack templates.  
\textit{Scam-V}~\cite{scamv} uses model-based relational testing and random test generation to discover undocumented cache-based leaks in ARM CPUs. 
In contrast, \Gemfuzzer{} focuses on \textit{pre}-silicon testing of speculation countermeasures early at design time, much before they are deployed in CPUs.


\smallskip
\noindent\textbf{RTL Fuzzing.}
Several works propose design-time testing of CPU designs at Register-Transfer Level (RTL).
\textit{WhisperFuzz}~\cite{borkar2024whisperfuzz} detects timing side channels in RTL using coverage-driven fuzzing, whereas \textit{SpecDoctor}~\cite{specdoctor} detects transient execution vulnerabilities in RTL using attack templates.
Neither work detects a wide variety of speculative leaks like \Gemfuzzer{}:
WhisperFuzz can only test for constant-time execution, and it is inapplicable to test most secure speculation countermeasures~\cite{STT,invisispec,cleanupspec,speclfb}; 
SpecDoctor tests only one type of leakage, where a secret in memory is exposed in speculation, and it lacks an extensible leakage model like \tool{}.
Other tools~\cite{hur2021difuzzrtl, kande2022thehuzz,cascade} that test CPU designs at RTL using golden reference models only detect functional bugs.
%
%
In contrast to RTL-based tools, \Gemfuzzer{} tests secure speculation countermeasures at the earliest stage of \uarch{} design, when features are being prototyped in simulators.
This allows computer architects to detect and address potential leaks in \uarch{} countermeasures early on.

\smallskip
\noindent
\textbf{Formal Verification.}
These approaches reason about and prove the absence of leaks in hardware designs.

\textit{Micro-architectural Tools.}
%
\textit{Checkmate}~\cite{checkmate} uses \uarch{} happens before graphs to generate security litmus tests.
%
%
\textit{Pensieve}~\cite{pensieve} uses bounded model checking to reason about the security of countermeasures~\cite{invisispec,ghostminion}  expressed as high-level \uarch{} models.
%
%
Both approaches require advanced modeling capabilities and considerable human effort to formalize the models in custom DSLs, which makes them impractical for design-phase testing. 
%
Their scalability is also limited: Checkmate has only been applied to simple in-order CPU models and attacks with up to 7 instructions, while Pensieve only analyzes programs up to 9 simulation cycles in a few hours.
In contrast, \Gemfuzzer{} can work with complex out-of-order CPU models in \uarch{} simulators, without artificial limits on simulation time or attack programs.
It also requires minimal integration effort, only requiring the addition of trace modules before testing can begin with new defenses.

%
\textit{RTL-Based Tools.} 
%
UPEC~\cite{fadiheh2022exhaustive} proves the absence of leaks caused by transient instructions in RTL designs, whereas LeaVe~\cite{ccs2023_zilong} proves whether an RTL design satisfies a leakage contract.
%
These tools require implementations of defenses in RTL to test them.
In contrast, \tool{} can test defenses early on in \uarch{} simulators, even before the RTL is generated.

\section{Conclusion}

\tool{} is the first tool to enable the testing of countermeasures for speculative leaks in \uarch{} simulators. 
Designed to achieve high testing speed and efficacy, AMuLeT enables large-scale campaigns on four countermeasures, revealing 3 known and 6 unknown leakages in them, including for the first time, a vulnerability in the implementation of the recently proposed SpecLFB. 
With \tool{}, we \rev{enable designers of} future defenses \rev{to test their counter-measure} at design time, reducing the risk of insecure deployments.

\begin{acks}
\newcounter{thesponsor}
\setcounter{thesponsor}{0}

This work is supported by 
\stepcounter{thesponsor}the \grantsponsor{\arabic{thesponsor}}{Spanish Ministry of Science and Innovation}{https://www.ciencia.gob.es/} under \grantnum{\arabic{thesponsor}}{TED2021-132464B-I00 PRODIGY}; 
\stepcounter{thesponsor}the \grantsponsor{\arabic{thesponsor}}{Spanish Ministry of Science and Innovation}{https://www.ciencia.gob.es/} under the Ram\'on y Cajal grant \grantnum{\arabic{thesponsor}}{RYC2021-032614-I};
\stepcounter{thesponsor}the \grantsponsor{\arabic{thesponsor}}{Spanish Ministry of Science and Innovation}{https://www.ciencia.gob.es/} under \grantnum{\arabic{thesponsor}}{PID2022-142290OB-I00 ESPADA};
\stepcounter{thesponsor} the \grantsponsor{\arabic{thesponsor}}{Natural Sciences and Engineering Research Council of Canada (NSERC)}{} under \grantnum{\arabic{thesponsor}}{RGPIN-2023-04796};
\stepcounter{thesponsor} the \grantsponsor{\arabic{thesponsor}}{Natural Sciences and Engineering Research Council of Canada (NSERC)}{} under \grantnum{\arabic{thesponsor}}{RGPIN-2021-03729};
an NSERC Undergraduate Student Research Award (USRA);
an SFU Faculty Recruitment Grant; and gifts from Intel Corporation and Ampere Computing.
\end{acks}

%
%
%
%
%






\appendix
\section{Artifact Appendix}

\subsection{Abstract}

This appendix describes artifacts accompanying the \tool{} paper, which introduces a tool for automated, design-time testing of secure speculation countermeasures. The artifact consist of two parts: (1) the \tool{} framework, our tool customized for testing defenses in simulators, and (2) extensions to the Gem5 implementation of defenses that we test with \tool{}. We include instructions for running campaigns with \tool{} on speculative execution countermeasures such as InvisiSpec, CleanupSpec, STT, and SpecLFB, and reproducing the key results in the paper.

\subsection{Artifact check-list (meta-information)}

{\small
\begin{itemize}
  \item {\bf Algorithm: } Implements \tool{}'s fuzzing techniques.
  \item {\bf Compilation: } Automatically through dockerfiles.
  \item {\bf Run-time environment: } Requires Docker. We use linux environments inside Docker to run Amulet and gem5.
  \item {\bf Hardware: }$\sim$100 cores, 128GB RAM to run parallel test campaigns.
  \item {\bf Metrics: } Test Throughput, Avg. Detection Time.
  \item {\bf Output: } Table 5 (results of test campaigns).
  \item {\bf Experiments: } Instructions to run campaigns are provided in the README.
  \item {\bf How much disk space required (approximately)?: }30GB
  \item {\bf How much time is needed to complete experiments (approximately)?: }80 hours
  \item {\bf Publicly available?: }Yes
  \item {\bf Code licenses (if publicly available)?: } MIT
  \item {\bf Workflow automation framework used?: } Docker
  \item {\bf Archived?: } \url{https://doi.org/10.5281/zenodo.14847073}
\end{itemize}
}

\subsection{Description}

\subsubsection{How to access}

\begin{itemize}
    \item {\bf \tool{}'s git repository}: \blue{\url{https://github.com/sith-lab/amulet}}. You can clone this directly from Github. 
    \item {\bf Defenses' git repository}: \blue{\url{https://github.com/sith-lab/amulet-gem5}}. This contains the defenses we test using \tool{}. This is cloned directly during the runs.
\end{itemize}


\subsubsection{Hardware dependencies}
\begin{itemize}
\item x86-based Server CPU
\item At least 128GB RAM, 100 cores to run parallel tests
\end{itemize}

\smallskip
\subsubsection{Software dependencies}
\begin{itemize}
\item \textbf{Docker}: We run all our tests inside docker containers. 
\item Each of our defenses have different software dependencies. So we provide Dockerfiles which automatically set up the required software dependencies. 

\end{itemize}



\subsection{Installation}

\begin{enumerate}
    \item Clone the GitHub repository:
\begin{tcolorbox}[colback=gray!10, colframe=black, boxrule=0.5pt, sharp corners, width=1\linewidth, arc=0mm, top=1pt, bottom=1pt, left=5pt]
\texttt{\footnotesize \$ git clone https://github.com/sith-lab/amulet.git}
\end{tcolorbox}

\item Run the artifact:

\begin{tcolorbox}[colback=gray!10, colframe=black, boxrule=0.5pt, sharp corners, width=1\linewidth, arc=0mm, top=1pt, bottom=1pt, left=5pt]
\texttt{\footnotesize \$ cd amulet ; ./run\_artifact.sh ;}
\end{tcolorbox}

This will run the campaigns for each of our defenses in \textbf{Table 5}, run InvisiSpec with reduced cache configurations to produce \textbf{Table 6}, and run Baseline (no defense) with different trace formats for \textbf{Table 4}.

\end{enumerate}


\subsection{Evaluation and expected results}
You should be able to reproduce Tables 4, 5, and 6, and discover violations in each defense. The runtimes and average detection time can vary based on differences in hardware capabilities and the exact campaign configuration run. 

\subsection{Experiment customization}
To reduce run times, one can decrease the number of test programs (default 200) as fewer test programs shorten run time. Fewer parallel campaigns (default 100) can be used on systems with fewer cores and memory. 








\bibliographystyle{plain}
\bibliography{references}
\end{document}